\documentclass[11pt]{amsart}
\usepackage{amsfonts,amssymb,multicol}

\def\func{\mathsf{F}}

\makeatletter
 \@addtoreset{equation}{section}
\makeatother

\def\double#1{#1,#1}

\def\tilde{\widetilde}

\def\theell{\mathsf{l}}

\newcommand{\cH}{\mathcal{H}}

\def\Relaxed{\raisebox{.5pt}{\rule{5pt}{6pt}}}
\def\Verma{\Large$\bullet$}
\def\TVerma{{\Large $\circ$}}

\newfont\sgb{cmmib8} 

\newcommand{\N}[1]{N\!=\!#1}
\newcommand{\SL}[1]{s\ell(#1)}
\newcommand{\tSL}[1]{\widehat{s\ell}(#1)}
\newcommand{\SSL}[2]{s\ell(#1|#2)}
\newcommand{\tSSL}[2]{\widehat{s\ell}(#1|#2)}

\def\NPB{Nucl.\ Phys.\ B}
\def\PLB{Phys.\ Lett.\ B}

\def\IJMPA{Int.\ J.\ Mod.\ Phys.\ A}

\newcommand\ellch{\mathsf{l}_{\mathrm{ch}}}

\newcommand{\half}{\frac{1}{2}}

\newcommand{\oA}{\mathbb{A}}
\newcommand{\oC}{\mathbb{C}}
\newcommand{\oN}{\mathbb{N}}

\newcommand{\oZ}{\mathbb{Z}}

\newcommand{\vt}[3]{\vartheta_{#1}(#2,#3)}

\newtheorem{Thm}{Theorem}[section]
\newtheorem{Lemma}[Thm]{Lemma}

\theoremstyle{definition}
\newtheorem{Dfn}[Thm]{Definition}

\theoremstyle{remark}
\newtheorem{Rem}[Thm]{Remark}
\newtheorem*{notation}{Notation}

\def\cC{\mathcal{C}}

\def\cG{\mathcal{G}}
\def\cH{\mathcal{H}}

\def\cL{\mathcal{L}}

\def\cO{\mathcal{O}}

\def\cQ{\mathcal{Q}}

\def\cT{\mathcal{T}}

\def\smm{\mathfrak}
\def\mM{{\smm{M}}}  
\def\mV{{\smm{V}}}  
\def\mU{{\smm{U}}}  
\def\mAs{{\smm{I}}} 
\def\mAn{{\smm{J}}} 
\def\mAm{{\smm{P}}} 
\def\mC{{\smm{C}}}  
\def\mD{{\smm{D}}}  
\def\mL{{\smm{L}}}  
\def\mK{{\smm{K}}}  
\def\smM{{\smm{M}}}
\def\smV{{\smm{V}}}
\def\mN{{\smm{N}}}  

\def\mm{\mathcal}
\def\mF{\mm{F}}

\def\mI{\mm{I}}

\newcommand{\charsl}[2]{{\chi_{
{{\phantom{h}\kern-3pt #2}}}^{\phantom{y}\kern-3pt #1}
}}  
\newcommand{\charn}[2]{{\omega_{
{{\phantom{h}\kern-3pt #2}}}^{\phantom{y}\kern-3pt #1}
}} 
\newcommand{\bcharsl}[2]{{\overline{\chi}_{
{{\phantom{h}\kern-3pt #2}}}^{\phantom{y}\kern-3pt #1}
}}
\newcommand{\bcharn}[2]{{\overline{\omega}_{
{{\phantom{h}\kern-3pt #2}}}^{\phantom{y}\kern-3pt #1}
}}

\newcommand{\shpv}[2]{\bigl(#1\,,\,#2\bigr)}
\newcommand{\shpvn}[2]{\bigl(#1\,,\,#2\bigr)}

\newcommand{\ket}[1]{\mathchoice{%
    {\left|{#1}\right\rangle}}{|{#1}\rangle}{|{#1}\rangle}{|{#1}\rangle}}
\newcommand{\ketSL}[1]{\left|{#1}\right\rangle_{\SL2}}

\newcommand{\kettop}[1]{\left|{#1}\right\rangle_{\mathrm{top}}}

\newcommand{\ketGH}[1]{\left|{#1}\right\rangle_{\mathrm{gh}}}

\def\ctop{\mathsf{c}}
\def\Ctop{\mathsf{C}}

\def\jtop{\,\mathsf{j}}
\def\htop{\,\mathsf{h}}

\newcommand{\res}{\mathop{\mathrm{res}}\limits}
\newcommand{\Tr}{\mathop{\mathrm{Tr}}^{\phantom{y}}\nolimits}
\newcommand{\spfsl}[1]{\mathop{{\mathsf{U}_{#1}}}}
\newcommand{\chevsl}{\mathop{\mathsf{T}}}
\newcommand{\achevsl}{\mathop{\mathsf{T'}}}

\newcommand{\spfn}[1]{\mathop{\mathsf{\bar U}_{#1}}}
\newcommand{\chevn}{\mathop\mathsf{\bar T}}
\newcommand{\achevn}{\mathop\mathsf{\bar T'}}
\newcommand{\simtimesr}
{\mathrel{{\times}\kern-2.6pt\raisebox{1.2pt}{\mbox{\tiny $|$}}}}
\newcommand{\simtimesl}
{\mathrel{\raisebox{1.2pt}{\mbox{\tiny $|$}}\kern-2.6pt{\times}}}

\def\emt{energy-momentum tensor}
\def\hw{highest-weight}

\def\Autsl{\mathsf{Aut}}
\def\Autn{\mathsf{Aut}}

\def\q{p'}

\def\tensor{\otimes}
\renewcommand{\d}{\partial}

\begin{document}
\hfuzz=1pt
\addtolength{\baselineskip}{4pt}



\title[Resolutions and Characters of $\N2$ Representations]{\hfill{
    \lowercase{\tt hep-th/9805179}}\\[12pt]
  Resolutions and Characters of Irreducible Representations of the
  $\N2$ Superconformal Algebra}

\author{B.~L.~Feigin}
\address{Landau Institute for Theoretical Physics, Russian
  Academy of Sciences}
\author{A.~M.~Semikhatov}
\address{Tamm Theory Division, Lebedev Physics Institute,
  Russian Academy of Sciences}
\author{V.~A.~Sirota}
\author{I.~Yu.~Tipunin}

\begin{abstract}
  We evaluate characters of irreducible representations of the $\N2$
  supersymmetric extension of the Virasoro algebra. We do so by
  deriving the BGG-resolution of the admissible $\N2$ representations
  and also a new ``$3,5,7,.\,.\,.$''-resolution in terms of twisted
  massive Verma modules.  We analyse how the characters behave under
  the automorphisms of the algebra, whose most significant part is the
  spectral flow transformations. The possibility to express the
  characters in terms of theta functions is determined by their
  behaviour under the spectral flow. We also derive the identity
  expressing every $\tSL2$ character as a linear combination of
  spectral-flow transformed $\N2$ characters; this identity involves a
  finite number of $\N2$ characters in the case of unitary
  representations.  Conversely, we find an integral representation for
  the admissible $\N2$ characters as contour integrals of admissible
  $\tSL2$ characters.
\end{abstract}

\maketitle

\thispagestyle{empty}
\flushcolumns

\setcounter{tocdepth}{3}

\vspace*{-36pt}

\begin{center}
  \parbox{.95\textwidth}{
    \begin{multicols}{2}
      {\footnotesize
        \tableofcontents}
    \end{multicols}
    }
\end{center}

\addtolength{\parskip}{2pt}

\section{Introduction} We consider representations of the $\N2$
supersymmetric extension of the Virasoro algebra~\cite{[Ade]} and of
the affine Lie algebra $\tSL2$ with the emphasis on the relation
between the two representation theories~\cite{[FST]}. \ We construct
the resolutions that allow us to evaluate characters of the admissible
$\N2$ representations and then show how these characters are related
to the admissible $\tSL2$ characters~\cite{[KW0]}.

\addtolength{\parskip}{2pt}

Expressions for some (unitary) $\N2$ characters were first proposed
in~\cite{[Dob],[M],[Kir]} (in a closed form---in terms of
theta-functions---in~\cite{[M],[RY]}), with the modular properties of
the unitary representation characters given in~\cite{[RY],[KW]}. \ It
was noticed later~on~\cite{[Doerr2]} that the embedding structure of
$\N2$ Verma modules is not as had previously been assumed.  This might
have invalidated the derivation of characters, since the structure of
the resolution used to derive the characters is obviously sensitive to
the embedding structure.

The structure of submodules in $\N2$ Verma modules was described
in~\cite{[ST4]} (or, in the equivalent language of relaxed $\tSL2$
modules, in~\cite{[FST]}), which has allowed the complete
classification of the $\N2$ embedding diagrams~\cite{[SSi]}. \ There
are two different types of Verma-like modules, called the {\it
  massive\/} and the {\it topological\/}\;($\equiv$chiral) ones, of
which the latter may appear as submodules of the former, but not vice
versa; more precisely, it is the {\it twisted\/}\footnote{In this
  paper, {\sl twist\/} and {\sl twisted\/} refer to the spectral-flow
  transform, to be defined below.} topological Verma modules that
appear as submodules.
The embedding structure~\cite{[SSi]} of massive $\N2$ Verma modules is
indeed more complicated than, e.g., in the well-known $\tSL2$
Verma-module case~\cite{[RCW],[Mal]}.\footnote{Among the recent
  findings related to the $\N2$ superconformal algebra, note also the
  calculation of semi-infinite $\N2$ cohomology, i.e., the cohomology
  of the critical $\N2$ string~\cite{[JL]}.}

However, the admissible $\N2$ representations that we mostly
concentrate on in this paper are the quotients of massive Verma
modules such that at least one {\it charged singular
  vector\/}~\cite{[BFK]} is necessarily quotiened away. A crucial fact
is then that the quotient of a massive Verma module over the submodule
generated from a charged singular vector is a twisted topological
Verma module~\cite{[ST3],[ST4]}. \ Therefore, the admissible $\N2$
representations are also the quotients of twisted topological Verma
modules. Now, the embedding structure of topological Verma modules is
{\it equivalent\/} to the embedding structure of $\tSL2$ Verma
modules; therefore, the resolutions are isomorphic and, thus, the
evaluation of the $\N2$ characters is rather straightforward; in
particular, it is not affected by the superfluous subsingular vectors,
which for the $\N2$ algebra are an artifact of a restricted choice of
vectors from which to generate the submodules (see~\cite{[ST4]} and
references therein).

This equivalence between the embedding structures of $\tSL2$ Verma
modules and topological $\N2$ Verma modules follows from the
fact~\cite{[FST]} that the categories of $\tSL2$ and $\N2$
representations are equivalent modulo the spectral flows.  Categories
(roughly, collections of {\it objects}, some of which are connected by
arrows---{\it morphisms}) can be related by functors (``mappings''
that send objects into objects and morphisms, into morphisms). Two
categories are called equivalent if there is a functor $\func$ between
them and the inverse functor $\func^{-1}$ (such that
$\func^{-1}\,\func$ takes every object into an isomorphic object); a
well-known example is that of (finite-dimensional) Lie algebras and
connected simply-connected Lie groups.
The $\tSL2$ and $\N2$ representation categories are equivalent modulo
the spectral flow; this can be formalised by introducing chains of
spectral-flow-transformed modules (on which the equivalence is strict,
see~\cite{[FST]}) or by considering two representations ``isomorphic''
if they are related by the spectral flow (which is {\it not\/} an
isomorphism in the usual sense!).

The $\tSL2$ and $\N2$ spectral flows are known
from~\cite{[TheBook],[FST]} (see also~\cite{[BH]}) and \cite{[SS]},
respectively.  More generally, given the basic symmetry algebra (of,
e.g., a conformal model) and its {\it automorphism\/} group, the full
space of states of the theory can be taken to include the modules
subjected to the automorphisms (unless these give rise to equivalent
representations). For the algebras dealt with in this paper,
the most significant part of the automorphisms is given by the
spectral flow transformations (for short, {\it twist}, particularly
when applied to representations).  In the familiar case of a positive
integer $\tSL2$ level $k$ (see, e.g.,~\cite{[W-Gep]} for applications
to the WZW model), adding the twisted representations does not give
anything new because the unitary (integrable) representations are
invariant under the spectral flow with even transformation parameters,
while the spectral flow with an odd parameter maps one unitary
representation into another.  The situation is different for rational
$k$: in that case, the theory {\it is\/} extended by adding the
twisted representations for all~$\theta\in\mathbb Z$.  On the $\N2$
side, the significance of the spectral flow is already seen for the
unitary representations, since it is needed for enumerating all the
$(k+1)(k+2)/2$ unitary representations; also, submodules of the
topological Verma modules are necessarily the {\it twisted\/}
topological Verma modules.

The characters must carry a representation of the automorphism group.
As regards the spectral flow transform, this requirement
(Eqs.~\eqref{spectral-sl2-general} and \eqref{spectral-n2-general} for
the $\tSL2$ and $\N2$ characters, respectively) is very useful in
deriving certain properties of characters (in this respect, it is
similar to the modular `covariance' condition; note that the modular
and the spectral flow transformations are not completely independent,
they are combined by taking a semidirect product). On the $\tSL2$
side, the general transformation properties are illustrated by the
known characters, while on the $\N2$ side, the behaviour under the
spectral flow is crucial as regards rewriting the characters in terms
of theta-functions.

There is a subtlety, partly of a terminological nature.  When dealing
with characters, one usually defines them as formal Laurent series; in
some cases, such a series determines a holomorphic or meromorphic
function, or a distribution. In the cases we are interested in
(particularly for the $\tSL2$ representations), the series may
converge once the variables belong to some complex domain, in which
case the character would determine a meromorphic function. This
function may, however, have different series expansions in different
domains, which in general correspond to characters of different
modules (such as those obtained by the action of the automorphism
group). We, thus, have sometimes to distinguish between the
character defined as a series and the function to which this series
converges inside a certain domain (but which can be continued outside
of it).  To stress this point, we will sometimes refer to the
meromorphic function representing the character as the {\it character
  function}.

Some character functions may be invariant under the spectral flow.
This is the case with the unitary $\N2$ and $\tSL2$ representations,
which are invariant under the spectral flow transformations with the
parameter $\theta=k+2$ and $\theta=2$, respectively.  This shows up in
the quasiperiodicity of the theta functions through which these
character functions are expressed.\footnote{On the other hand, the
  admissible $\tSL2$ characters are invariant under the spectral flow
  with $\theta=2\q$ (where $k+2=\frac{p}{\q}$), whereas the
  representations are not, which is in accordance with the fact that a
  chosen character function gives characters of different modules (in
  the present case, of spectral-flow-transformed modules) when
  expanded in different domains.} \ On the unitary representations,
the orbits of the $\N2$ spectral flow have length $k+2$, with
inequivalent unitary representation filling a half of the boxes in the
table on p.~\pageref{boxes} (into which the Ka\v c table is included
as the first column).  At the same time, the $\tSL2$ spectral flow
orbits on unitary representations are of length~2; in accordance with
the equivalence theorem, however, both algebras have {\it the same\/}
number $[\frac{k+2}{2}]$ of the {\it equivalence classes\/} of unitary
\hbox{representations modulo the spectral flow}.

\medskip

Including the twisted (spectral-flow-transformed) representations in
both the $\tSL2$ and $\N2$ theories is crucial in order to derive the
essential equivalence of the corresponding representation
categories~\cite{[FST]}. \ An immediate consequence of the
equivalence, as we see in Sec.~{\it\ref{sec:identities}}, is that the
$\tSL2$ characters are given by infinite sums of the $\N2$ characters
transformed by the spectral flow with all $\theta\in\oZ$:
\begin{equation*}
    \charsl{}{j,t}(z,q)\,\vt{1,0}{\tfrac{z}{y}}{q}=
    \sum_{\theta\in\oZ}
    \charn{}{-\frac{2j}{t},t}(y^{-1}\,q^{-\theta}, q)\,
    y^{\theta-\frac{2}{t}j}\,z^{j-\theta}\,
    q^{\frac{\theta^2-\theta}{2} +
      \frac{j^2 + j - 2 j \theta}{t} + \frac{1}{8}},
\end{equation*}
which can also be viewed as a ``branching'' of $\tSL2$ characters into
$\N2$ ones (or, a ``sumrule'' for the $\N2$ characters). This
applies to {\it any\/} $\N2$ representation; in particular, in the
simple case where these are taken as the topological Verma modules,
this formula becomes the ``truly remarkable identity''
from~\cite{[KW]} (where it was derived using different
representation-theory considerations).

More identities are obtained by taking other representations. For the
unitary ones, the above-mentioned invariance under the spectral flow
with $\theta=k+2$ on the $\N2$ side results in that the identity
contains only a finite number (precisely $k+2$) of different $\N2$
characters, their linear combination giving a unitary $\tSL2$
character (see~\eqref{unitary-identity}).  For the admissible
representations, on the other hand, the identity contains an infinite
number of $\N2$ characters.  The reason is that neither the admissible
$\N2$ representations themselves nor their character functions are
periodic under the spectral flow. This also implies that the
admissible $\N2$ characters cannot be {\it algebraically\/} expressed
in terms of the theta functions.  However, there is an integral
representation for the admissible $\N2$ character functions
$\charn{\mAn}{r,s,p,\q}$ through those of the corresponding (hence,
admissible) $\tSL2$ representations,
\begin{multline*}
  \charn{\mAn}{r,s,p,\q}(y,q)\,y^{\frac{\q}{p}(r-1)+1-s}=
  \frac{1}{2\pi i}\oint
  \frac{d z}{z}\,\vt{1,0}{zy}{q}\,
  \frac{q^{-\frac{1}{4}p\q}}{
    \vartheta_{1,1}(z, q)}\times{}\\
  {}\times\left(
    \vartheta_{1,0}(z^p\,q^{-p\q  +  r \q  - (s-1) p}, q^{2p\q})
    -z^{-r}\,q^{r(s-1)}\,
    \vartheta_{1,0}(z^p\,q^{-p\q  -  r \q  - (s-1) p}, q^{2p\q})
  \right),
\end{multline*}
see~\eqref{integral-admissible} for more details; in fact, a similar
representation exists for any $\N2$ character,
see~\eqref{integral-general}.

\medskip

The means to derive characters of irreducible representations is
provided by {\it resolutions\/} of these representations in terms of
Verma modules. As has already been said, the $\N2$ BGG-resolution in
terms of only (twisted) topological Verma modules is of the same
structure as the well-known $\tSL2$ resolution~\cite{[BGG]}. However,
using the massive $\N2$ Verma modules opens up other possibilities.

To arrive at a resolution in terms of massive Verma modules, we
start with the $\N2$ embedding diagrams~\cite{[SSi]}; these contain,
generically, modules of both types, the twisted topological and the
massive ones. We then rewrite the embedding diagram as an {\it
  exact\/} sequence (in particular, the mappings between modules
acquire kernels, thereby no longer being {\it embeddings\/}); this
`re-interpretation' is not completely automatic for sufficiently
complicated embedding diagrams (as, e.g., the III$^0_+(2,{-}{+})$ one,
which we explicitly consider in what follows).  Next, we replace
every topological Verma module with its resolution in terms of massive
Verma modules.  Examining the resulting mappings, we can read off the
resolution of a new type, with the number of modules growing from term
to term as 3, 5, 7,\ldots, see~\eqref{massive-resolution}.

All these structures exist also in the $\tSL2$ guise: as explained
in~\cite{[FST],[SSi]}, the $\tSL2\leftrightarrow\N2$ equivalence
implies the following ``dictionary'' between the objects of the two
representation theories:
\begin{center}
  \renewcommand{\arraystretch}{0}
  \begin{tabular}{|p{200pt}|p{220pt}|}
    \hline\multicolumn{2}{|c|}{\rule{0pt}{1.5pt}}\\\hline
    \strut $\tSL2$ & $\N2$\\
    \hline\multicolumn{2}{|c|}{\rule{0pt}{1.5pt}}\\\hline
    \strut(twisted) Verma module &
    twisted topological($\equiv${\it chiral\/}) Verma module\\\hline
    \strut singular vector in Verma module & topological singular
    vector \\
    \hline
    \strut (twisted) relaxed Verma module~\cite{[FST]} & (twisted)
    massive Verma module \\
    \hline
    \strut unitary representations ($k+1$ inequivalent representations
    for level $k\in\oN
$) & unitary
    representations [$(k+1)(k+2)/2$ inequivalent representations for
    central charge $\ctop=3(1-2/(k+2))$]\\
    \hline
    \strut periodicity 2 under the spectral flow for the unitary
    representations & periodicity $k+2$
    under the spectral flow for the unitary representations\\
    \hline
  \end{tabular}
\end{center}

\smallskip

\noindent
In what follows, we add to the list the admissible representations of
each algebra. However, we do not consider the relaxed $\tSL2$ Verma
modules here, although their $\N2$ counterpart will play a role in our
constructions.

\smallskip

In Sec.~\ref{sec:general}, we introduce the $\tSL2$ and $\N2$ algebras
and the respective spectral flows. In Sec.~\ref{sec:resolutions}, we
recall the BGG resolution of the $\tSL2$ admissible representations
and then derive the BGG resolution of the $\N2$ admissible
representations and a new resolution consisting entirely of massive
Verma modules. In Sec.~\ref{sec:irreducible}, we evaluate the
admissible representation characters and find how they behave under
the spectral flow (for each of the two algebras). We also consider in
some detail the unitary representations, which are interesting to us
here primarily because they are periodic under the spectral flow,
which has a number of consequences. In Sec.~\ref{sec:correspondence},
we relate the $\N2$ and $\tSL2$ characters using the equivalence of
categories, which implies ``sum rules'' and an integral representation
for the $\N2$ characters.  Appendix~\ref{app:theta} summarizes our
theta-function conventions.

\medskip
\addtolength{\parskip}{-4pt}

\begin{notation} We use the following notations for various
  modules:
  \begin{tabbing}
    (twisted) topological $\N2$ Verma modules,~~\=$\mV$,\qquad\qquad
    \=$\tSL2$ admissible representations,~~\=$\mAs$,\kill
    (twisted) topological $\N2$ Verma modules, \>$\mV$,
    \>$\tSL2$ Verma modules, \>$\mM$,\\
    (twisted) massive $\N2$ Verma modules, \>$\mU$,
    \mbox{}\\
    $\N2$ admissible representations, \>$\mAn$,
    \>$\tSL2$ admissible representations, \>$\mAs$,\\
    $\N2$ unitary representations, \>$\mK$,
    \>$\tSL2$ unitary representations, \>$\mL$.
  \end{tabbing}
  We also write $\charn{}{}$ and $\charsl{}{}$ for the $\N2$ and
  $\tSL2$ characters, respectively.
\end{notation}

\section{The algebras, their Verma modules, and spectral flows
  \label{sec:general}}
In this section, we introduce the $\tSL2$ and $\N2$ algebras and show
how the respective spectral flows act between representations and,
thus, act also on characters.

\subsection{The $\protect\tSL2$ side}
\subsubsection{The algebra and the spectral flow} The affine $\SL2$
algebra is defined by the commutation relations
\begin{align}
  {[}J^0_m,\,J^\pm_n]&={}\pm J^\pm_{m+n}\,,\qquad
  [J^0_m,\,J^0_n]={}\tfrac{K}{2}\,m\,\delta_{m+n,0}\,,
  \displaybreak[0]\\
  {[}J^+_m,\,J^-_n]&={}K\,m\,\delta_{m+n,0} + 2J^0_{m+n},
  \label{sl2modes}
\end{align}
with $K$ being the central element, whose eigenvalue will be denoted
by~$t-2$ with $t\in\oC\setminus\{0\}$.

For $\theta\in\oZ$, we have the automorphisms given by the spectral
flow transformations~\cite{[FST]} (a general definition can be found
in~\cite{[TheBook]})
\begin{equation}
  \spfsl\theta:\quad
  J^+_n\mapsto J^+_{n+\theta}\,,\qquad
  J^-_n\mapsto J^-_{n-\theta}\,,\qquad
  J^0_n\mapsto J^0_n+\tfrac{t-2}{2}\,\theta\delta_{n,0}\,.
  \label{spectral-sl2}
\end{equation}
We also have the involutive automorphism
\begin{equation}
  \chevsl:\quad
  J^+_n\mapsto J^-_{n}\,,\qquad
  J^-_n\mapsto J^+_{n}\,,\qquad
  J^0_n\mapsto -J^0_n\,.
\end{equation}
These satisfy
\begin{equation}
  \spfsl\theta\spfsl{\theta'}=\spfsl{\theta+\theta'}\,,\qquad
  \spfsl0=\chevsl{}^2=1\,,\qquad
  \spfsl\theta\chevsl=\chevsl\spfsl{-\theta}\,.
  \label{autrel}
\end{equation}
The automorphism group is thus the semidirect product $\Autsl=
\oZ\simtimesl \oZ_2$. It contains the affine Weyl group
$(2\oZ)\simtimesl \oZ_2$ as the {\it inner\/} automorphisms.  We also
have the involutive {\it anti\/}-automorphism
\begin{equation}
\achevsl:\quad
  J^+_n\mapsto J^-_{-n}\,,\qquad
  J^-_n\mapsto J^+_{-n}\,,\qquad
  J^0_n\mapsto J^0_{-n}\,
\label{antiautsl}
\end{equation}
that commutes with the group $\Autsl$.

The character $\charsl{\mC}{t}\equiv\charsl{\mC_t}{}$ of a module
$\mC_t$ with a definite value of the level $K=t-2$ is defined by
\begin{equation}\label{lyuboiverma}
  \charsl{\mC}{t}(z,q)=\Tr_{\mC_t}\left(
    q^{L_0}\,z^{J^0_0}
  \right),
\end{equation}
where $L_0$ is the zero mode of the Sugawara \emt. We assume here a
sesquilinear (Shapovalov) form such that for any $A\in\tSL2$ we have
$A^\dagger=\achevsl A$.
As to $q$ and $z$, these are complex variables and we assume $|q|<1$
in what follows. The series implied in~\eqref{lyuboiverma} would in
general converge only in some domain depending on the module under
consideration. One often divides $\charsl{}{t}(z,q)$ by $z^jq^\Delta$,
where $j$ and $\Delta$ are the charge and the Sugawara dimension of
the \hw{} vector, but we do not do it here.

Elements of the automorphism group~$\Autsl$ can be applied to any
$\tSL2$ module~$\mC$; we thus obtain the modules
$\spfsl\theta\mC=\mC_{;\theta}$ and~$\chevsl\mC$ (the former are
called the `{\sl twisted\/}' modules). Hence the automorphism group
also acts on character functions of $\tSL2$ modules:
\begin{equation}
  \spfsl\theta\charsl{\mC}{}(z,q)
  =\charsl{\spfsl\theta\mC}{}(z,q)\,,\qquad\chevsl\charsl{\mC}{}(z,q)=
  \charsl{\chevsl\mC}{}(z,q)\,.
\end{equation}
For a module with a definite level $K=t-2$, we write
$\charsl{\mC}{t;\theta}(z,q)$ for
$\spfsl{\theta}\charsl{\mC}{t}(z,q)$.
\begin{Lemma} Let $\mC\equiv\mC_t$ be an $\tSL2$ module with the
  level $K=t-2$. Then its character function transforms under the
  spectral flow transform and the involution as
  \begin{equation}\label{spectral-sl2-general}
    \charsl{\mC}{t;\theta}(z,q)=
    q^{\frac{t-2}{4}\theta^2}\,z^{-\frac{t-2}{2}\theta}\,
    \charsl{\mC}{t}(z\,q^{-\theta},q)\,,\qquad
    \chevsl\charsl{\mC}{t}(z,q)=\charsl{\mC}{t}(z^{-1},q)\,.
  \end{equation}
  If, further, the series for $\charsl{\mC}{t;\theta}(z,q)$ converges
  in an annulus $\oA$, the transformed character converges in the
  annulus
\begin{equation}
  \spfsl\theta\oA=\oA_\theta=
  \{z\bigm|
  z\,q^{-\theta}\in\oA\}\,,\qquad
  \chevsl\oA=\oA^{-1}=\{z\bigm|z^{-1}\in\oA\}.
\end{equation}
\end{Lemma}

\subsubsection{Verma modules and twisted Verma modules}
The action of the spectral flow on Verma modules gives {\it twisted
  Verma modules\/}~$\smM_{j,t;\theta}$ \cite{[FST]}, which are
described as follows.  For $\theta\in\oZ$, the twisted Verma module
$\smM_{j,t;\theta}$ is freely generated by $J^+_{\leq\theta-1}$,
$J^-_{\leq-\theta}$, and $J^0_{\leq-1}$ from a twisted highest-weight
vector $\ketSL{j,t;\theta}$ defined by 
\begin{equation}
  \begin{split}
    J^+_{\geq\theta}\,\ketSL{j,t;\theta}&=
    J^0_{\geq1}\,\ketSL{j,t;\theta}=
    J^-_{\geq-\theta+1}\,\ketSL{j,t;\theta}=0\,,\\
    \left(J^0_{0}+\tfrac{t-2}{2}\theta\right)\,\ketSL{j,t;\theta}&=
    j\,\ketSL{j,t;\theta}\,,
  \end{split}\quad\theta\in\oZ\,.
  \label{sl2higgeneral}
\end{equation}
The \hw{} vector $\ketSL{j,t;\theta}$ of the twisted Verma module has
the Sugawara dimension
\begin{equation}
  \Delta(j,t;\theta)=
  \tfrac{j^2+j}{t}-\theta j + \tfrac{t-2}{4}\theta^2\,.
\end{equation}
Setting $\theta=0$ in the above formulae gives the usual (`untwisted')
Verma modules. We define $\ketSL{j,t}=\ketSL{j,t;0}$, and similarly,
denote $\mM_{j,t}=\smM_{j,t;0}$.

\medskip

The character of a Verma module $\mM_{j,t}$ converges for $|q|<1$ and
$z\in\oA$,
\begin{equation*}
  \oA=\{1<|z|<|q|^{-1}\}\,,
\end{equation*}
where it can be summed to the character function
\begin{equation}\label{sl-Verma}
  \charsl{\mM}{j,t}(z,q)=
  \frac{q^{\frac{j^2+j}{t} + \frac{1}{8}}\,z^{j}}{
    \vartheta_{1,1}(z, q)}
\end{equation}
(see the Appendix for our conventions on theta functions).  To obtain
occupation numbers in each grade, one should expand~\eqref{sl-Verma}
assuming that $z\in\oA$.  As to the behaviour under the spectral flow,
the character of a twisted Verma module converges in
\begin{equation*}
  \oA_\theta=\{z\bigm|
  |q|^\theta<|z|<|q|^{\theta-1}\}\,,
\end{equation*}
where Eq.~\eqref{spectral-sl2-general} allows us to find
\begin{equation}
  \charsl{\mM}{j,t;\theta}(z,q)=
  (-1)^\theta\,q^{\theta^2\frac{t}{4}-\theta(j+\half)}\,
  z^{-\frac{t}{2}\theta}\,
  \charsl{\mM}{j,t}(z,q)\,.
\end{equation}
To obtain occupation numbers in each grade, this should be expanded
assuming that $z$ is inside the annulus~$\oA_\theta$. We also assume
$|q|<1$ everywhere in what follows.

\subsection{The $\N2$ side} We now introduce the $\N2$ superconformal
algebra and the corresponding analogues of the structures considered
above for~$\tSL2$. Note that the proper $\N2$ counterpart of the
$\tSL2$ Verma modules are the topological Verma modules considered in
Sec.~{\it\ref{sec:top-Verma}}, while the massive Verma modules
introduced in Sec.~{\it\ref{sec:masive}\/} are somewhat different
objects, whose $\tSL2$ analogues are not considered here.

\subsubsection{The algebra and its automorphisms} The $\N2$
superconformal algebra is taken in the basis where
\begin{alignat}{2}\label{topalgebra}
  {[}\cL_m,\cL_n]&=(m-n)\cL_{m+n}\,,&[\cH_m,\cH_n]&=
  \tfrac{\Ctop}{3}m\delta_{m+n,0}\,,\notag\displaybreak[0]\\
  {[}\cL_m,\cG_n]&=(m-n)\cG_{m+n}\,,&
  [\cH_m,\cG_n]&=\cG_{m+n}\,, \notag\displaybreak[0]\\
  {[}\cL_m,\cQ_n]&=-n\cQ_{m+n}\,,&
  [\cH_m,\cQ_n]&=-\cQ_{m+n}\,,\displaybreak[0]\\
  {[}\cL_m,\cH_n]&=
  -n\cH_{m+n}+\tfrac{\Ctop}{6}(m^2+m)\delta_{m+n,0}\,,\kern-80pt
  \notag\\
  \{\cG_m,\cQ_n\}&=
  2\cL_{m+n}-2n\cH_{m+n}+
  \tfrac{\Ctop}{3}(m^2+m)\delta_{m+n,0}\,,\kern-80pt\notag
\end{alignat}
and $m,~n\in\oZ$. In what follows, we do not distinguish between the
central element $\Ctop$ and its eigenvalue~$\ctop$, which we assume to
be $\ctop\neq3$ and parametrise as $\ctop=3(1-\frac{2}{t})$
with~$t\in\oC\setminus\{0\}$.

The automorphisms of the $\N2$ algebra consist of the spectral
flow~\cite{[SS]} and the involutive automorphism.  In the basis chosen
in~\eqref{topalgebra}, the spectral flow acts as
\begin{gather}
  \spfn{\theta}:
  \begin{aligned}
    \cL_n&\mapsto\cL_n+\theta\cH_n+\tfrac{\ctop}{6}(\theta^2+\theta)
    \delta_{n,0}\,,&\qquad{}
    \cH_n&\mapsto\cH_n+\tfrac{\ctop}{3}\theta\delta_{n,0}\,,\\
    \cQ_n&\mapsto\cQ_{n-\theta}\,,&{}\cG_n&\mapsto\cG_{n+\theta}\,,
  \end{aligned}
  \label{U}
\end{gather}
where $\theta\in\oZ$. The involutive automorphism is given by
\begin{equation}
  \chevn:
  \begin{aligned}
    \cG_n&\mapsto \cQ_{n}\,,\qquad& \cQ_n&\mapsto \cG_{n}\,,\\
    \cH_n&\mapsto -\cH_n-\tfrac{\ctop}{3}\delta_{n,0}\,,\quad&
    \cL_n&\mapsto \cL_n-n\cH_n\,.
  \end{aligned}
\end{equation}
As in the $\tSL2$ case, the operations $\spfn{\theta}$ and $\chevn$
satisfy \eqref{autrel} and thus constitute the group of automorphisms
$\Autn=\oZ\simtimesl\oZ_2$.  We also have an involutive {\it
  anti\/}-automorphism
\begin{equation}
  \achevn:
  \begin{aligned}
    \cG_n&\mapsto \cQ_{-n}\,,\qquad& \cQ_n&\mapsto \cG_{-n}\,,\\
    \cH_n&\mapsto \cH_{-n}\,,\quad&
    \cL_n&\mapsto \cL_{-n}+n\cH_{-n}\,,
  \end{aligned}
  \label{antiautn}
\end{equation}
which commutes with the group $\Autsl$.

We introduce the characters as
\begin{equation}
  \charn{\mD}{}(z,q) = \Tr_{\mD}\left(
    z^{\cH_0}\,q^{\cL_0}
  \right)
\end{equation}
and write $\charn{\mD}{t}(z,q)\equiv\charn{\mD_t}{}(z,q)$ for a module
$\mD_t$ with a definite central charge $\ctop=3(1-\frac{2}{t})$. We
assume here a sesquilinear form~\cite{[BFK]} such that for any element
$A$ from the $\N2$ algebra we have $A^\dagger=\achevn A$
or equivalently,
\begin{alignat}{2}\label{Shap-N2}
  \shpvn{\ket y}{\cG_n\ket z}&=
  \shpvn{\cQ_{-n}\ket y}{\ket z}\,,& \shpvn{\ket y}{\cQ_n\ket z}&=
  \shpvn{\cG_{-n}\ket y}{\ket z}\,,\\
  \shpvn{\ket y}{\cL_n\ket z}&=
  \shpvn{(\cL_{-n}+n\cH_{-n})\ket y}{\ket z}\,,\quad&
  \shpvn{\ket y}{\cH_n\ket z}&=
  \shpvn{\cH_{-n}\ket y}{\ket z}
\end{alignat}
for $\ket y$ and $\ket z$ from the module.

The modules are mapped under automorphisms, which we denote as
$\mD_{t;\theta}\equiv\spfn{\theta}\mD_t$ in the case of the spectral
flow. Therefore, the automorphisms act on characters, for which we
write $\charn{\mD}{t;\theta}\equiv\spfn{\theta}\charn{\mD}{t}$.
\begin{Lemma}
  Let $\mD\equiv\mD_{t}$ be an $\N2$ module with a definite central
  charge $\ctop=3(1-\frac{2}{t})$. Then the character function of the
  spectral-flow transformed module $\mD_{t;\theta}$ is
  \begin{equation}\label{spectral-n2-general}
    \charn{\mD}{t;\theta}(z,q) = z^{-\frac{\ctop}{3}\theta}\,
    q^{\frac{\ctop}{6}(\theta^2 - \theta)}\,
    \charn{\mD}{t}(z\,q^{-\theta},q)\,.
  \end{equation}
  For the involutive automorphism, similarly,
  \begin{equation}\label{invautcharn}
    \chevn\charn{\mD}{t}(z,q)=z^{-\frac{\ctop}{3}}\,
    \charn{\mD}{t}(z^{-1},q)\,.
  \end{equation}
\end{Lemma}
\begin{Rem}
  For a given $\charn{\mD}{t}(z,q)$, Eq.~\eqref{spectral-n2-general}
  also allows one to determine the characters transformed by the
  spectral flow with {\it half-integral\/}~$\theta$, in particular
  $\theta=\half$, thereby recovering the ``NS sector'' of the algebra.
  We will not repeat this point and keep on working with the
  characters that directly pertain to the algebra written in the
  basis~\eqref{topalgebra}.

  Note also that in the $\N2$ context, it is not necessary to
  distinguish between characters (defined via formal series) and
  character functions, since the series converge for $z\in\oC$ and,
  thus, the characters are holomorphic functions.
\end{Rem}

\subsubsection{Topological Verma modules\label{sec:top-Verma}} We now
consider the class of $\N2$ Verma modules that we call {\it
  topological\/}\footnote{{\it chiral\/}, in a different set of
  conventions, see, e.g.,~\cite{[LVW]}.} Verma modules
following~\cite{[FST],[ST4]}. \ We give the definitions applicable to
the twisted case as well.  First, for a fixed $ \theta\in\oZ$, we
define the {\it twisted topological \hw{} vector\/}
$\kettop{h,t;\theta}$ to satisfy the annihilation conditions
\begin{equation}
  \cQ_{-\theta+m}\kettop{h,t;\theta}=
  \cG_{\theta+m}\kettop{h,t;\theta}=
  \cL_{m+1}\kettop{h,t;\theta}=
  \cH_{m+1}\kettop{h,t;\theta}=0\,,\quad m\in\oN_0\,,
  \label{annihiltop}
\end{equation}
with the following eigenvalues of the Cartan generators (where the
second equation follows from the annihilation conditions):
\begin{align}
    (\cH_0+\tfrac{\ctop}{3}\theta)\,\kettop{h,t;\theta}&=
    h\,\kettop{h,t;\theta}\,,\\
    (\cL_0+\theta\cH_0+\tfrac{\ctop}{6}(\theta^2+\theta))
    \,\kettop{h,t;\theta}&=0\,.
\end{align}
\begin{Dfn}
  The {\it twisted topological Verma module\/} $\mV_{h,t;\theta}$ is
  the module freely generated from the topological \hw{} vector
  $\kettop{h,t;\theta}$ by $\cQ_{\leq-1-\theta}$,
  $\cG_{\leq-1+\theta}$, $\cL_{\leq-1}$, and $\cH_{\leq-1}$.
\end{Dfn}

The twisted topological \hw{} vectors $\kettop{h,t;\theta}$ are
defined in accordance with the action of $\spfn{}$, so that
$\spfn{\theta'}\kettop{h,t;\theta}=\kettop{h,t;\theta+\theta'}$.  We
write $\kettop{h,t}\equiv \kettop{h,t;0}$ in the `untwisted' case of
$\theta=0$ and also denote by $\mV_{h,t}\equiv\mV_{h,t;0}$ the
untwisted module.\footnote{The untwisted modules are in a certain
  sense more ``rare'' in the $\N2$ context than in the $\tSL2$ one,
  because submodules of a topological Verma module are always the {\it
    twisted\/} topological Verma modules~\cite{[ST4]}.}

The character of the topological Verma module~$\mV_{h,t}$ converges
for $|q|<1$ and $z\in\oC$ and can then be summed to
\begin{equation}
  \charn{\mV}{h,t}(z,q)=
  z^{h}\frac{\vartheta_{1,0}(z,q)}{
    \displaystyle(1 + z^{-1})\,\eta(q)^3}\,.
  \label{topchar}
\end{equation}
The character function of a twisted topological Verma module, thus,
reads as
\begin{equation}
  \charn{\mV}{h,t;\theta}(z,q)=
  z^{h + \frac{2\theta}{t}}\,
  q^{-h\theta-\frac{\theta^2-\theta}{t}}\,
  \frac{\vartheta_{1,0}(z,q)}{
    \displaystyle(1 + z^{-1}q^\theta)\,\eta(q)^3}\,.
\end{equation}

\subsubsection{Massive Verma modules over the $\N2$
  algebra\label{sec:masive}} A different class of Verma-like $\N2$
modules are defined as follows~\cite{[ST4]}.
\begin{Dfn}
  A twisted massive Verma module $\mU_{h,\ell,t;\theta}$ is freely
  generated from a twisted {\it massive \hw{} vector\/}
  $\ket{h,\ell,t;\theta}$ by the generators
  \begin{equation}
    \cL_{-m}\,,~m\in\oN\,,\qquad \cH_{-m}\,,~m\in\oN\,,\qquad
    \cQ_{-\theta-m}\,,~m\in\oN_0\,,\qquad \cG_{\theta-m}\,,~m\in\oN\,.
    \label{verma}
  \end{equation}
  The massive \hw{} vector $\ket{h,\ell,t;\theta}$ satisfies the
  following conditions:
  \begin{gather}\label{masshw}
    \cQ_{m+1-\theta}\,\ket{h,\ell,t;\theta}=
    \cG_{m+\theta}\, \ket{h,\ell,t;\theta}=
    \cL_{m+1}\,\ket{h,\ell,t;\theta}=
    \cH_{m+1}\,\ket{h,\ell,t;\theta}=0\,,~m\in\oN_0\,,\\
    \begin{aligned}
      (\cH_0 + \tfrac{\ctop}{3}\theta)\,\ket{h,\ell,t;\theta}&=
      h\,\ket{h,\ell,t;\theta}\,,\\
      (\cL_0 + \theta\cH_0 + \tfrac{\ctop}{6}(\theta^2 +
      \theta))\,\ket{h,\ell,t;\theta} &=
      \ell\,\ket{h,\ell,t;\theta}\,.
    \end{aligned}
  \end{gather}
\end{Dfn}
\noindent
We also write $\ket{h,\ell,t}=\ket{h,\ell,t;0}$ and
$\mU_{h,\ell,t}=\mU_{h,\ell,t;0}$.

The character of the twisted massive Verma module is given by
\begin{equation}\label{massive-character}
  \charn{\mU}{h,\ell,t;\theta}(z,q)
  =z^{h+\frac{2}{t}\theta}\, q^{\ell-\theta h -
    \frac{\theta^2-\theta}{t}}\,
  \frac{\vartheta_{1,0}(z,q)}{\eta(q)^3}\,.
\end{equation}

A charged singular vector occurs in $\mU_{h,\ell,t}$
whenever~\cite{[BFK]} $\ell=\ellch(n,h,t)\equiv -n(h-\frac{n+1}{t})$,
$n\in\oZ$, and reads as~\cite{[ST3],[ST4]}
\begin{equation}
  \ket{E(n,h,t)}_{\mathrm{ch}}=\left\{
    \begin{aligned}
      \cQ_{-n}\,\ldots\,\cQ_0\,\ket{h,\ellch(n,h,t),t}\,,
      &\quad&n\geq0\,,\\
      \cG_{n}\,\ldots\,\cG_{-1}\,\ket{h,\ellch(n,h,t),t}\,,&&
      n\leq-1\,.
    \end{aligned}\right.
  \label{ECh}
\end{equation}
This state on the extremal diagram~\cite{[ST4]}
satisfies the {\it twisted\/} topological \hw{} conditions with the
twist~$\theta=n$, the submodule generated from
$\ket{E(n,h,t)}_{\mathrm{ch}}$ being the twisted topological Verma
module $\mV_{h-\frac{2n}{t}-1,t;n}$ if $n\geq0$ and
$\mV_{h-\frac{2n}{t},t;n}$ if $n\leq-1$.

\section{Some $\protect\tSL2$ and $\N2$
  resolutions\label{sec:resolutions}} For the rest of the paper, we
fix two coprime positive integers $p$ and $\q$ and parametrise
$t=\frac{p}{\q}$.

For positive rational~$t$, we consider (twisted) topological Verma
modules with dominant \hw s.  Such a module is not embedded into
another (twisted) topological Verma module and admits infinitely many
singular vectors.  We define the {\it admissible $\N2$
  representations\/} to be the (irreducible) quotients of these Verma
modules over the maximal submodules.

The functors from~\cite{[FST]} (which we review in Sec.~{\it
  \ref{subsec:equivalence}\/}) relate the $\N2$ and $\tSL2$ admissible
representations to each other (note that we extend ``admissible
representations'' to mean also all the integral twists of the
``standard'' admissible $\tSL2$ representations~\cite{[KW0]}).
Therefore, the embedding diagrams of the $\N2$ topological Verma
modules are isomorphic to the embedding diagrams of the $\tSL2$ Verma
modules and, thus, the corresponding resolutions are isomorphic as
well (as objects in the category of exact sequences). In addition to
the thus obtained BGG-resolution, the admissible $\N2$ representations
possess ``massive'' resolutions constructed in the category of twisted
massive Verma modules.

\subsection{The standard BGG resolution\label{sec:BGG}}
\subsubsection{The $\protect\tSL2$ side\label{BGG-sl2}} As is
well-known~\cite{[TheBook]}, the admissible representation
$\mAs_{r,s,p,\q}$ is the quotient over the maximal submodule of the
Verma module $\mM_{\jtop(r,s,\frac{p}{\q}),\frac{p}{\q}}$, where
\begin{equation}\label{adm-sl2}
  2\jtop(r,s,\tfrac{p}{\q}) + 1 = r - (s-1)\tfrac{p}{\q}\,,\qquad
  1\leq r\leq p-1\,,\quad 1\leq s\leq \q\,.
\end{equation}
The BGG-resolution for the admissible $\tSL2$ representations is the
exact sequence
\begin{equation}\label{resolution}
  \ldots \to \mN_-(m) \oplus \mN_+(m) \to
  \ldots\to\mN_-(2) \oplus \mN_+(2) \to \mN_-(1) \oplus
  \mN_+(1) \to \mN \to \mI \to 0\,,
\end{equation}
where $\mI\approx\mAs_{r,s,p,\q}$ is the irreducible representation in
question,
$\mN\approx\mM_{\frac{r-1}{2}-\frac{p}{2\q}(s-1),\frac{p}{\q}}$ is the
Verma module whose quotient is $\mAs_{r,s,p,\q}$, and
\begin{equation}\label{legend}
  \begin{split}
    \mN_+(2m)&\approx
    \mM_{\frac{r-1}{2}-\frac{p}{2\q}(s-1)-pm,\frac{p}{\q}}\,,\\
    \mN_+(2m-1)&\approx
    \mM_{-\frac{r+1}{2}-\frac{p}{2\q}(s-1)-p(m-1),
      \frac{p}{\q}}\,,
  \end{split}
  \quad
  \begin{split}
    \mN_-(2m)&\approx
    \mM_{\frac{r-1}{2}-\frac{p}{2\q}(s-1)+pm,\frac{p}{\q}}\,,\\
    \mN_-(2m-1)&\approx
    \mM_{-\frac{r+1}{2}-\frac{p}{2\q}(s-1)+pm,\frac{p}{\q}}\,,
  \end{split}
  \quad m\in\oN\,.
\end{equation}
are the Verma modules embedded into $\mN$.

In order to enumerate all the admissible representations (including,
in accordance with our conventions, the spectral-flow-transformed
ones), we should take all the twists $\mAs_{r,s,p,\q;\theta}$,
$\theta\in\oZ$, of $\mAs_{r,s,p,\q}$. This then results in that every
module in \eqref{resolution}--\eqref{legend} acquires the additional
twist~${}_{;\theta}$.

\subsubsection{The $\N2$ side\label{BGG-N2}} All the $\N2$ admissible
representations (which we denote as $\mAn_{r,s,p,\q;\theta}$) can be
obtained as the quotients of {\it twisted\/} topological Verma modules
$\mV_{\htop(r,s,\frac{p}{\q}),\frac{p}{\q};\theta}$, where
$\theta\in\oZ$ and
\begin{equation}
  \htop(r,s,t)=-\tfrac{r-1}{t}+s-1,\quad
  t=\tfrac{p}{\q}, \quad 1\leq r\leq p-1, \quad 1\leq s\leq \q\,.
\end{equation}
\begin{Thm}\label{thm:resolution-N}
  There is an exact sequence of $\N2$ representations given
  by~\eqref{resolution}, where $\mI\approx\mAn_{r,s,p,\q;\theta}$ is
  an admissible representation, $\mN\approx\mV_{-\frac{\q}{p}(r-1) +
    s-1,\frac{p}{\q};\theta}$ is a twisted topological Verma module,
  and, for $m\in\oN$,
  \begin{align*}
        \mN_+(2m)&\approx
        \mV_{-\frac{\q}{p}(r-1) + s-1 + 2\q
          m,\frac{p}{\q};\theta-mp}\,,\displaybreak[3]\\
        \mN_+(2m-1)&\approx
        \mV_{\frac{\q}{p}(r+1) + s-1 + 2\q
          (m-1),\frac{p}{\q}; \theta -r-(m-1)p}\,,\\
        \mN_-(2m)&\approx
        \mV_{-\frac{\q}{p}(r-1) + s-1 - 2\q m,\frac{p}{\q};
          \theta+mp}\,,\\
        \mN_-(2m-1)&\approx
        \mV_{\frac{\q}{p}(r+1) + s-1 - 2\q
          m,\frac{p}{\q}; \theta+mp-r}\,,
  \end{align*}
  are twisted topological Verma modules.
\end{Thm}
\begin{proof}[Sketch of the Proof.]
  The only difference from the $\tSL2$ case is that the resolution
  consists of {\it twisted\/} topological Verma modules with different
  twists.  Although the submodules in a twisted topological Verma
  module $\mV$ appear simultaneously with submodules in the
  corresponding $\tSL2$ Verma module $\mM$, yet the submodules in
  $\mV$ are necessarily twisted (all with different twists, in
  general):
\begin{equation}
  \label{compare}
  \unitlength=.5pt
  \begin{picture}(600,200)
    \put(0,20){
      \bezier{400}(0,0)(50,120)(118,166)
      \bezier{400}(118,166)(190,120)(240,0)
      \bezier{200}(20,0)(40,50)(70,85)
      \bezier{300}(70,85)(170,140)(230,0)
      \put(240,100){$\Longleftrightarrow$}
      \put(300,160){\line(1,0){180}}
      \put(480,160){\line(1,-1){120}}
      \put(310,100){\line(1,0){100}}
      \put(410,100){\line(1,-1){100}}
      }
    \put(80,0){$\N2$}\put(370,0){$\tSL2$}
  \end{picture}
\end{equation}
The equivalence statement does not fix the twists, and these have to
be found separately. They are determined by the properties of
topological singular vectors.  To a singular vector $\ket{{\rm
    MFF}(r,s,t)}^\pm$, $r,s\in\oN$ (see~\cite{[FST]} for the
definition in the current notations) in the $\tSL2$ Verma module
$\mM_{j,t}$, there corresponds the {\it topological singular
  vector\/}~\cite{[ST4]} $\ket{E(r,s,t)}^\pm\in \mV_{-2j/t,t}$ that
satisfies the $\theta=\mp r$-twisted topological \hw{} conditions:
\begin{equation}
  \cQ_{\geq\pm r}\ket{E(r,s,t)}^\pm=
  \cG_{\geq\mp r}\ket{E(r,s,t)}^\pm=
  \cL_{\geq1}\ket{E(r,s,t)}^\pm=
  \cH_{\geq1}\ket{E(r,s,t)}^\pm=0
  \label{twistedtophw}
\end{equation}
(these states are at the grades where the extremal diagram of the
submodule shown in~\eqref{compare} has a ``cusp'').  Thus, the
submodule generated from
$\ket{E^\pm(r,s,t)}^{\pm;\theta}\in\smV_{h,t;\theta}$ (where
${}^{;\theta}$ denotes the spectral flow transform of the singular
vector) is the twisted topological Verma module $\smV_{h\pm
  r\frac{2}{t},t;\theta\mp r}$, hence the twists $m p$ and $m p-r$ of
the modules in the resolution.
\end{proof}

The twisted topological Verma submodule is {\it freely\/} generated
from the topological singular vector.  At the same time, these
singular vectors generate {\it maximal\/} submodules, which is crucial
for the resolution to have precisely the form~\eqref{resolution} (in
particular, the spurious subsingular vectors have no effect on
it~\cite{[ST4]}, as we also see from the correspondence with the
$\tSL2$ case).

\subsection{A ``massive'' resolution of the admissible
  representations\label{sec:massive}}
Another resolution for admissible $\N2$ representations can be arrived
at by considering the {\it massive\/} Verma module whose quotient is
the admissible representation in question. Our strategy is as follows.
We take from~\cite{[SSi]} the embedding diagram of the appropriate
massive Verma module~$\mU$ and resolve every twisted topological Verma
module appearing in the diagram in terms of (twisted) massive Verma
modules.  In this way, we arrive at a diagram that represents all the
mappings (which are no longer embeddings!) into~$\mU$ of twisted
massive Verma modules, then all the mappings {\it into\/} these
modules, and so on.  {}From this {\it diagram of morphisms}, we then
read off the resolution.

The admissible $\N2$ representation $\mAn_{r,s,p,\q;\theta}$ is the
quotient of the massive Verma module $\mU_{h,\ell,\frac{p}{p'}}$ with
the \hw{} parameters
\begin{equation}\label{take-massive}
  \ell= -\theta  (h - \tfrac{\theta +1}{t})\,,\qquad
  h = \left\{\begin{array}{ll}
      -\tfrac{r-1}{t} + s + \tfrac 2t \theta, & \theta\leq-1\,,\\
      -\tfrac{r-1}{t} + s-1  + \tfrac 2t \theta, & \theta \ge 0\,,
    \end{array}
  \right.\quad t=\tfrac{p}{\q}\,,\quad
  \begin{array}{l}1\le r\le p-1\,,\\
    1 \le s \le \q
  \end{array}
\end{equation}
over the maximal submodule.  The embedding diagrams of the Verma
modules with parameters~\eqref{take-massive} are different for $s=1$
and $2\leq s\leq \q$. We now concentrate on the case of $s=1$ (where
the embedding diagram is more complicated).  Then the massive Verma
module is included into the following diagram of morphisms of twisted
massive Verma modules:
\begin{equation}\label{admisss1}
  \unitlength=.8pt
  \begin{picture}(400,140)
    \put(0,-110){
       \put(320,245){$\scriptstyle\cdots$}
       \put(293,245){$\scriptstyle3$}
       \put(265,245){$\scriptstyle2$}
       \put(237,245){$\scriptstyle1$}
       \put(209,245){$\scriptstyle0$}
       \put(176,245){$\scriptstyle-1$}
       \put(148,245){$\scriptstyle-2$}
       \put(120,245){$\scriptstyle-3$}
       \put(96,245){$\scriptstyle\cdots$}
       \put(70,217){$\scriptstyle0$}
       \put(70,190.3){$\scriptstyle1$}
       \put(70,163.6){$\scriptstyle2$}
       \put(70,136.9){$\scriptstyle3$}
       \put(70,110.7){$\scriptstyle\cdots$}
      {\linethickness{.1pt}
        \multiput(80,220)(0,-26.7){5}{\line(1,0){250}}
        \multiput(100.5,240)(28,0){9}{\line(0,-1){130}}
        }
      \multiput(0,0)(0,-26.5){4}{
        \put(209,216){{\Relaxed}}
        \put(236.5,196){\vector(-1,1){20}}
        \put(187.5,196){\vector(1,1){20}}
        \put(212.3,197){\vector(0,1){19}}
        }
      \multiput(-28.0,-26.5)(0,-26.5){3}{
        \put(209,216){{\Relaxed}}
        \put(236.5,196){\vector(-1,1){20}}
        \put(187.5,196){\vector(1,1){20}}
        \put(212.3,197){\vector(0,1){19}}
        }
      \multiput(28.0,-26.5)(0,-26.5){3}{
        \put(209,216){{\Relaxed}}
        \put(236.5,196){\vector(-1,1){20}}
        \put(187.5,196){\vector(1,1){20}}
        \put(212.3,197){\vector(0,1){19}}
        }
      \multiput(-56.0,-53)(0,-26.5){2}{
        \put(209,216){{\Relaxed}}
        \put(236.5,196){\vector(-1,1){20}}
        \put(187.5,196){\vector(1,1){20}}
        \put(212.3,197){\vector(0,1){19}}
        }
      \multiput(56.2,-53)(0,-26.5){2}{
        \put(209,216){{\Relaxed}}
        \put(236.5,196){\vector(-1,1){20}}
        \put(187.5,196){\vector(1,1){20}}
        \put(212.3,197){\vector(0,1){19}}
        }
      \put(84.3,-79.5){
        \put(209,216){{\Relaxed}}
        \put(236.5,196){\vector(-1,1){20}}
        \put(187.5,196){\vector(1,1){20}}
        \put(212.3,197){\vector(0,1){19}}
        }
      \put(-84.0,-79.5){
        \put(209,216){{\Relaxed}}
        \put(236.5,196){\vector(-1,1){20}}
        \put(187.5,196){\vector(1,1){20}}
        \put(212.3,197){\vector(0,1){19}}
        }
      }
  \end{picture}
\end{equation}
We label the module at the intersection of the $a$-th row and the
$b$-th column by~$\mU(a,b)$ and set $\theta=0$, with the nonzero
$\theta$ to be restored in the end. We also temporarily omit
$t=\frac{p}{\q}$ from the notations for modules, writing
$\mU_{h,\ell;\theta}$ for $\mU_{h,\ell,t;\theta}$ (and, accordingly,
$\mU_{h,\ell}$ for $\mU_{h,\ell,t}$).
\begin{Thm}\label{thm:s=1}
  For $s=1$, there is the exact sequence
  \begin{equation}\label{massive-resolution}
    \cdots\longrightarrow\bigoplus_{b=-a}^{a}\mU(a,b)\longrightarrow
    \cdots\longrightarrow\mU(1,1)\oplus\mU(1,0)\oplus\mU(1,-1)
    \longrightarrow\mU(0,0)\longrightarrow\mAn_{r,1,p,\q;0}
    \longrightarrow0\,,
  \end{equation}
  where
  \begin{equation}\label{massive-modules1}
    \bigoplus_{i=-2n}^{2n}\mU(2n,i)=
    \begin{array}[t]{l}
      \bigoplus\limits_{j=1}^n
      \mU_{\double{1-\frac{\q}{p}(r+1) + 2\q j + 2(n-j)};1-pj}\\
      \quad{}\oplus
      \bigoplus\limits_{j=0}^{n-1}
      \mU_{\double{\frac{\q}{p}(r-1) + 2 \q j + 2(n-j)};1-r-p j}
      \oplus
      \mU_{\frac{\q}{p}(1-r)-2\q n,0;n p}\\
      \qquad{}\oplus\bigoplus\limits_{j=1}^n
      \mU_{\frac{\q}{p}(r+1)-1-2\q j - 2(n-j),0;pj-r}
      \oplus\bigoplus\limits_{j=0}^{n-1}
      \mU_{\frac{\q}{p}(1-r) - 2\q j - 2(n-j),0;pj}
    \end{array}
  \end{equation}
  and
  \begin{equation}\label{massive-modules2}
    \bigoplus_{i=-2n-1}^{2n+1}\mU(2n+1,i)=
    \begin{array}[t]{l}
      \bigoplus\limits_{j=0}^n
      \mU_{\double{\frac{\q}{p}(r-1) + 1 + 2\q j + 2(n-j)};1-r-pj}\\
      \quad{}\oplus
      \bigoplus\limits_{j=1}^{n}
      \mU_{\double{2-\frac{\q}{p}(r+1) + 2 \q j + 2(n-j)};1-p j}\\
      \qquad{}\oplus
      \mU_{\frac{\q}{p}(r+1)-2\q(n+1),0;(n+1)p-r}
      \oplus\bigoplus\limits_{j=0}^n
      \mU_{\frac{\q}{p}(1-r)-1-2\q j - 2(n-j),0;pj}\\
      \quad\qquad{}\oplus\bigoplus\limits_{j=1}^{n}
      \mU_{\frac{\q}{p}(1+r) - 2\q j - 2(1+n-j),0;pj-r}\,.
    \end{array}
  \end{equation}
\end{Thm}

The rest of this subsection is devoted to proving this.

The starting point is the relevant embedding diagram
from~\cite{[SSi]}, which is given by case III$^0_+(2,{-}{+})$.  To
describe it, we also borrow some notations from~\cite{[SSi]}. \ The
arrows in the {\it embedding\/} diagrams are drawn from the `parent'
module to its submodule. Horizontal arrows denote embeddings onto {\it
  dense $\cQ$- or $\cG$-descendants\/} (see~\cite{[ST4],[SSi]}). The
(twisted) massive Verma submodules are denoted by {\Relaxed}, while
{\Verma} and {\TVerma} denote twisted topological Verma
submodules.\footnote{The difference between the {\Verma} and {\TVerma}
  modules is that the former are generated from a twisted topological
  \hw{} state $\ket{e'}$ satisfying twisted topological \hw{}
  conditions with the twist parameter $\theta'$ and
  $\cH_0\ket{e'}=(h_0-\theta') \ket{e'}$, with $h_0$ being the
  eigenvalue of $\cH_0$ on the \hw{} vector of~$\mU$, while the latter
  are generated from $\ket{e'}$ satisfying twisted topological \hw{}
  conditions with the twist parameter $\theta'$ and
  $\cH_0\ket{e'}=(h_0-\theta'-1) \ket{e'}$.} \ Now the
III$^0_+(2,{-}{+})$ embedding diagram reads~as
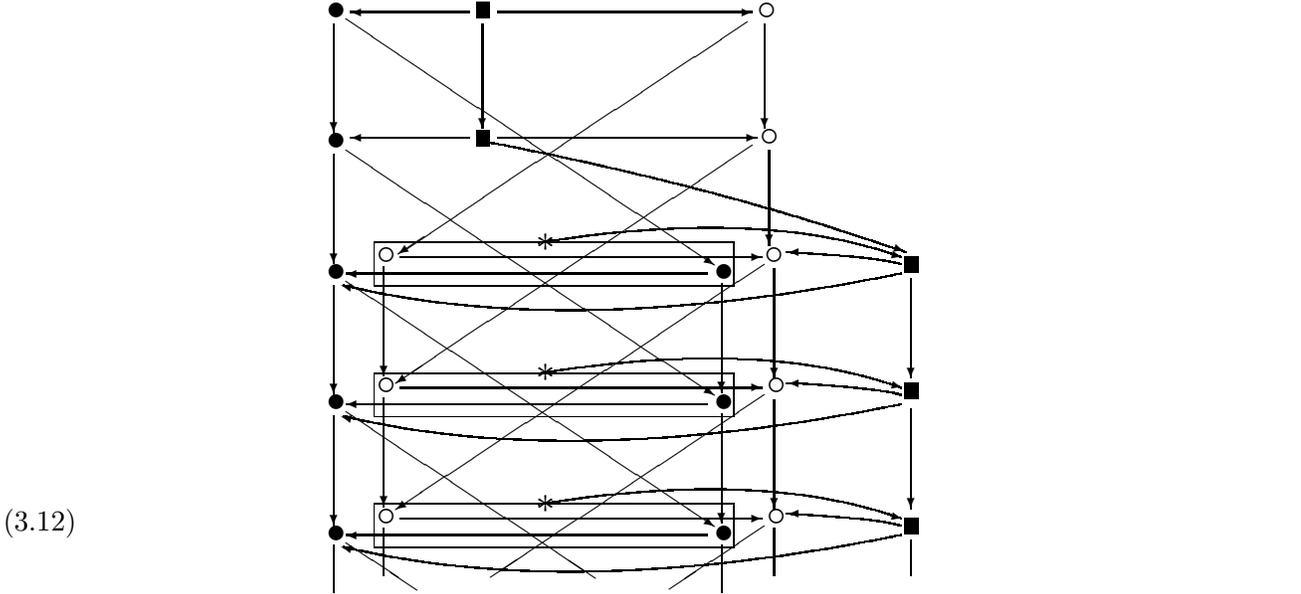
\begin{equation}
  \label{d:IIIpm0(2,mp)}
  \unitlength=0.9pt
  \begin{picture}(500,225)
    \put(50,0){
    \bezier{400}(116,163)(210,145)(290,117)
    \put(288,118){\vector(3,-1){2}}
    \put(292.5,106){\vector(0,-1){42}}
    \put(135,117.6){\Large$\ast$}
    \bezier{300}(136,121)(230,136)(288,115)
    \put(287,115.5){\vector(4,-1){2}}
    \bezier{500}(288,108)(150,80)(54,103)
    \put(55,102.5){\vector(-4,1){2}}
    \bezier{100}(288,112)(278,115)(240,117)
    \put(243,117){\vector(-1,0){2}}
    \put(0,-55){
    \put(292.5,106){\vector(0,-1){42}}
      \put(135,117.6){\Large$\ast$}
      \bezier{300}(136,121)(230,136)(288,115)
      \put(287,115.5){\vector(4,-1){2}}
      \bezier{500}(288,108)(150,80)(54,103)
      \put(55,102.5){\vector(-4,1){2}}
      \bezier{100}(288,112)(278,115)(240,117)
      \put(243,117){\vector(-1,0){2}}
      }
    \put(0,-110){
    \put(292.5,106){\line(0,-1){15}}
      \put(135,117.6){\Large$\ast$}
      \bezier{300}(136,121)(230,136)(288,115)
      \put(287,115.5){\vector(4,-1){2}}
      \bezier{500}(288,108)(150,80)(54,103)
      \put(55,102.5){\vector(-4,1){2}}
      \bezier{100}(288,112)(278,115)(240,117)
      \put(243,117){\vector(-1,0){2}}
      }
    {\linethickness{.2pt}
      \put(67.00,103.00){\framebox(151.00,18.00)[cc]{\mbox{}}}
      }
    \put(0,-55){
      {\linethickness{.2pt}
        \put(67.00,103.00){\framebox(151.00,18.00)[cc]{\mbox{}}}
        }
      }
    \put(0,-110){
      {\linethickness{.2pt}
        \put(67.00,103.00){\framebox(151.00,18.00)[cc]{\mbox{}}}
        }
      }
    \put(210,125){
      \put(-163,90){\Verma}
      \put(-155,90){\vector(3,-2){155}}
      \put(-160,88){\vector(0,-1){46}}
      \put(0,-55){
        \put(-163,90){\Verma}
        \put(-155,90){\vector(3,-2){155}}
        \put(-160,88){\vector(0,-1){46}}
        }
      \put(0,-110){
        \put(-163,90){\Verma}\put(-3,93){\vector(-1,0){152}}
        \put(0,90){\Verma}
        \put(3,89){\vector(0,-1){46}}
        \put(-155,90){\vector(3,-2){155}}
        \put(-160,88){\vector(0,-1){46}}
        }
      \put(0,-165){
        \put(-163,90){\Verma}\put(-3,93){\vector(-1,0){152}}
        \put(0,90){\Verma}
        \put(3,89){\vector(0,-1){46}}
        \put(-155,90){\line(3,-2){105}}
        \put(-160,88){\vector(0,-1){46}}
        }
      \put(0,-220){
        \put(-163,90){\Verma}\put(-3,93){\vector(-1,0){152}}
        \put(0,90){\Verma}
        \put(3,89){\line(0,-1){20}}
        \put(-155,90){\line(3,-2){30}}
        \put(-160,89){\line(0,-1){20}}
        }
      }
    \put(210,125){
      \put(-100,90){\Relaxed}
      \put(-97.5,88){\vector(0,-1){44}}
      \put(-103,93){\vector(-1,0){50}} \put(-91,93){\vector(1,0){107}}
      \put(-100,36){\Relaxed}
      \put(-103,40){\vector(-1,0){50}} \put(-91,40){\vector(1,0){109}}
      }
    \put(75,132){
      \put(153,83){\TVerma}
      \put(149,82){\vector(-3,-2){147}}
      \put(156,81){\vector(0,-1){44}}
      \put(0,-55){
        \put(154,85){\TVerma}
        \put(151,85){\vector(-3,-2){150}}
        \put(158,83){\vector(0,-1){40}}
        }
      \put(0,-110){
        \put(156,90){\TVerma}\put(3,93){\vector(1,0){152}}
        \put(-7,90){\TVerma}
        \put(-4,89){\vector(0,-1){46}}
        \put(156,90){\vector(-3,-2){155}}
        \put(160,88){\vector(0,-1){46}}
        }
      \put(0,-165){
        \put(157,90){\TVerma}\put(3,93){\vector(1,0){152}}
        \put(-7,90){\TVerma}
        \put(-4,89){\vector(0,-1){46}}
        \put(156,90){\line(-3,-2){115}}
        \put(160,88){\vector(0,-1){46}}
        }
      \put(0,-220){
        \put(157,90){\TVerma}\put(3,93){\vector(1,0){152}}
        \put(-7,90){\TVerma}
        \put(-4,89){\line(0,-1){20}}
        \put(156,90){\line(-3,-2){40}}
        \put(160,89){\line(0,-1){20}}
        }
      }
   \put(290, 108){\Relaxed}
    \put(290, 55){\Relaxed}
    \put(290, -2){\Relaxed}
}
  \end{picture}
\end{equation}

\bigskip

\bigskip

\medskip

\noindent
The frame around the pairs of {\Verma} and {\TVerma} modules
represents the direct sum of these modules. Accordingly, an arrow
drawn from that frame (symbolized by a $\ast$) to the corresponding
twisted massive Verma module indicates that the latter is embedded
into the direct sum. On the other hand, the twisted massive Verma
module has two submodules associated with charged singular vectors, as
shown by the arrows drawn {\it from\/} the massive Verma module.

At the top level in~\eqref{d:IIIpm0(2,mp)}, we have the massive Verma
module $\mU$ with its two twisted topological Verma submodules $\mV_-$
and $\mV_+$ generated from the charged singular vectors. Each of these
submodules has the standard lattice of submodules represented by a
``double-braid'' embedding diagram; the corresponding submodules are,
again, twisted topological Verma modules. We denote by $\mV_-$ the
{\Verma}-submodule generated from $\ket{E(n,h,t)}_{\mathrm{ch}}$ with
$n\leq-1$, and by $\mV_+$, the {\TVerma}-module generated from
$\ket{E(m,h,t)}_{\mathrm{ch}}$ with~$m\geq0$.  Then the submodules of
the {\Verma} type are those embedded into $\mV_-$ (with $\mV_-$ itself
being the top one), while all those of the {\TVerma} type are embedded
into (and including)~$\mV_+$.

We now need to introduce notations for the remaining modules
in~\eqref{d:IIIpm0(2,mp)}. \ At the next level, we have a (twisted)
massive Verma module $\mU^1$ with its two submodules embedded via
charged singular vectors, $\mV_-^1$ and $\mV_+^1$ (such that,
$\mV_\mp^1\subset\mV_\mp$, in fact $\mV^1_\mp=\mU^1\cap\mV_\mp$).
One level down, we have a (twisted) massive Verma module $\mU^2$ and
{\it four\/} twisted topological Verma modules. The {\Verma} ones are
denoted by $\mV^2_-$ and $\tilde\mV^2_-$ so that
$\mV^2_-\subset\tilde\mV^2_-$, and the {\TVerma} ones are
$\tilde\mV^2_+\supset\mV^2_+$.  We have~\cite{[ST4],[SSi]} \
$\mU^2\cap\tilde\mV^2_-=\mV^2_-$ and $\mU^2\cap\tilde\mV^2_+=\mV^2_+$.
This pattern then repeats at every embedding level: we have the
{\Verma}-modules $\mV^i_-\subset\tilde\mV^i_-$ and the {\TVerma}-ones
$\tilde\mV^i_+\supset\mV^i_+$, such that
\begin{equation}\label{modules}
\mU^i\cap\tilde\mV^i_-=\mV^i_-, \quad\mU^i\cap\tilde\mV^i_+=\mV^i_+,
\qquad i\geq2\,.
\end{equation}

A crucial step leading from the embedding
diagram~\eqref{d:IIIpm0(2,mp)} to the resolution is the fact that the
massive Verma module $\mU^i$ is embedded~\cite{[SSi]} into the direct
sum of the corresponding $\tilde\mV^i_-$ and $\tilde\mV^i_+$,
\begin{equation}
  \mU^i\hookrightarrow\tilde\mV^i_-\oplus\tilde\mV^i_+\,,
  \qquad i\geq2\,,
\end{equation}
with $\mU^i\cap\,\tilde\mV^i_-=\mV^{i-1}_-$ and
$\mU^i\cap\tilde\mV^i_+=\mV^{i-1}_+$. This leads to the following
Lemma.  We first note that the irreducible representation
$\mAn\equiv\mAn_{r,1,p,\q}$ is given by $0 \to \mV_- + \mU^1 + \mV_+
\to\mU\to\mAn\to0$ (where the sum is {\it not\/} direct) with the
embedding given by, obviously,
\begin{equation}\label{1st-embedding}
  \mV_- + \mU^1 + \mV_+ \to \mU\,,\qquad
  (x_-,u^1,x_+)\mapsto x_- + u^1 + x_+.
\end{equation}
Now,
\begin{Lemma}\label{lemma:mixed} There is the exact sequence
  \begin{equation}\label{mixed}
    \ldots
    \to \mV^2_- \oplus \mU^3 \oplus \mV^2_+
    \to \mV^1_- \oplus \mU^2 \oplus \mV^1_+
    \to \mV_- \oplus \mU^1 \oplus \mV_+
    \to \mU \to\mAn\to 0\,.
  \end{equation}
\end{Lemma}
\begin{proof}[Sketch of the Proof.]
  To see what these mappings are, consider, e.g., $\mV^1_- \oplus
  \mU^2 \oplus \mV^1_+ \to \mV_- \oplus \mU^1 \oplus \mV_+$. Using the
  embedding
  $\mU^2\hookrightarrow\widetilde\mV^2_-\oplus\widetilde\mV^2_+$, we
  represent an element $u^2\in\mU^2$ as $\widetilde x^2_- + \widetilde
  x^2_+$, then the mapping $\mV^1_- \oplus \mU^2 \oplus \mV^1_+ \to
  \mV_- \oplus \mU^1 \oplus \mV_+$ given by
  \begin{equation}\label{2nd-embedding}
    (x^1_-,\; \widetilde x^2_- + \widetilde x^2_+,\; x^1_+)\mapsto
    (x^1_- - \widetilde x^2_-,\;
    \widetilde x^2_- + \widetilde x^2_+ - x^1_- - x^1_+,\;
    x^1_+ - \widetilde x^2_+)
  \end{equation}
  spans the kernel of~$\mV^1_- \oplus \mU^2 \oplus \mV^1_+ \to \mV_-
  \oplus \mU^1 \oplus \mV_+$.  It is also instructive to consider the
  next mapping, $\mV^2_- \oplus \mU^3 \oplus \mV^2_+ \to \mV^1_-
  \oplus \mU^2 \oplus \mV^1_+$. As before, we use the fact that
  $\mU^3\hookrightarrow\widetilde\mV^3_- \oplus \widetilde\mV^3_+$,
  then
  \begin{equation}\label{3rd-embedding}
    (x^2_-,\; \widetilde x^3_- + \widetilde x^3_+,\; x^3_+)\mapsto
    (x^2_- + \widetilde x^3_-,\;
    \widetilde x^3_- + \widetilde x^3_+ + x^2_- + x^2_+,\;
    x^3_+ + \widetilde x^3_+)\,.
  \end{equation}
  {\it Because of the embeddings\/}
  $\widetilde\mV^3_\mp\hookrightarrow\widetilde\mV^2_\mp$ (which by
  themselves are parts of the embedding diagrams of $\mV_-$ and
  $\mV_+$, respectively), the element $\widetilde x^3_- + \widetilde
  x^3_+\in\mU^2$ can naturally be viewed as an element of
  $\widetilde\mV^2_-\oplus\widetilde\mV^2_+$, whence we see that the
  composition of \eqref{3rd-embedding} and \eqref{2nd-embedding}
  vanishes.

  Continuing in the same way, we complete the proof by taking into
  account that the quotient with respect to either $\mV_-$ or $\mV_+$
  gives a twisted topological Verma module.  For example, taking the
  quotient $\mV=\mU/\mV_+$, we obtain the same irreducible
  representation simply as $0 \to \mV_- + \mV^1 \to \mV \to\mAn\to 0$,
  where $\mV^1=\mU^1/\mV^1_+$. This then has the BGG-resolution
  \begin{equation}
    \ldots
    \to \mV^2_- \oplus \widetilde\mV^3_-
    \to \mV^1_- \oplus \widetilde\mV^2_-
    \to \mV_- \oplus \mV^1 \to \mV \to\mAn \to 0
  \end{equation}
  (note the occurrence of the $\mV^i_-$ modules, see~\eqref{modules}
  and the text before that formula).  A crucial point is that
  \begin{equation*}
    (\widetilde\mV^{i}_-+\widetilde\mV^{i}_+)/
    ((\widetilde\mV^{i}_-+\widetilde\mV^{i}_+)\cap\mV_+)=
    (\widetilde\mV^{i}_-+\widetilde\mV^{i}_+)/\widetilde\mV^{i}_+=
    \widetilde\mV^{i}_-,\qquad i\geq2\,.
  \end{equation*}
  The resolutions are therefore related as shown in the following
  exact commutative diagram:
  \begin{equation}\label{square1}
    \begin{array}{ccccccccccccl}
      \ldots
      &\!\!\longrightarrow\!\!& 0
      &\!\!\longrightarrow\!\!& 0
      &\!\!\longrightarrow\!\!& 0
      &\!\!\longrightarrow\!\!& 0 &\!\!\longrightarrow\!\!& 0 &
      \longrightarrow& 0\\
      {}&{}&\Bigm\downarrow{}&{}&\Bigm\downarrow{}&{}&\Bigm\downarrow
      {}&{}&\Bigm\downarrow{}&{}&\Bigm\downarrow\\
      \ldots
      &\!\!\longrightarrow\!& \mV^3_- \oplus \mV^2_+
      &\!\!\longrightarrow\!& \mV^2_+ \oplus \mV^1_+
      &\!\!\longrightarrow\!& \mV^1_+ \oplus \mV_+
      &\!\!\longrightarrow    & \mV_+ &\!\!\longrightarrow\!\!& 0 &
      \longrightarrow& 0\\
      {}&{}&\Bigm\downarrow{}&{}&\Bigm\downarrow{}&{}&\Bigm\downarrow
      {}&{}&\Bigm\downarrow{}&{}&\Bigm\downarrow\\
      \ldots
      &\!\!\longrightarrow\!\!& \mV^2_- \oplus \mU^3 \oplus \mV^2_+
      &\!\!\longrightarrow\!\!& \mV^1_- \oplus \mU^2 \oplus \mV^1_+
      &\!\!\longrightarrow\!\!& \mV_- \oplus \mU^1 \oplus \mV_+
      &\!\!\longrightarrow\!\!& \mU &\longrightarrow&
      \mAn&\!\!\longrightarrow\!\!& 0\\
      {}&{}&\Bigm\downarrow{}&{}&\Bigm\downarrow{}&{}&\Bigm\downarrow
      {}&{}&\Bigm\downarrow{}&{}&\Bigm\|\\
      \ldots&\!\!\longrightarrow\!\!&
      \kern-2pt\mV_-^2 \oplus\widetilde\mV_-^3\kern-2pt
      &\!\!\longrightarrow\!\!& \kern-2pt \mV_-^1\oplus\widetilde
      \mV_-^2\kern-2pt
      &\!\!\longrightarrow\!\!& \kern-2pt\mV_-\oplus\mV^1\kern-2pt &
      \!\!\longrightarrow\!\!& \mV &
      \!\!\longrightarrow\!\!& \mAn&\longrightarrow& 0\\
      {}&{}&\Bigm\downarrow{}&{}&\Bigm\downarrow{}&{}&\Bigm\downarrow
      {}&{}&\Bigm\downarrow{}&{}&\Bigm\downarrow\\
      \ldots
      &\!\!\longrightarrow\!\!& 0
      &\!\!\longrightarrow\!\!& 0
      &\!\!\longrightarrow\!\!& 0
      &\!\!\longrightarrow\!\!& 0 &\!\!\longrightarrow\!\!& 0 &\longrightarrow& 0,
    \end{array}
  \end{equation}
  Applying here the Lemma on Five Homomorphisms shows~\eqref{mixed}
  (the overall spectral transform by $\theta$ goes through the above
  argument by simply adding $\theta$ to the twist of every
  module).
\end{proof}

A straightforward analysis shows that the exact sequence~\eqref{mixed}
actually reads as (omitting $t=\frac{p}{\q}$ from the notations for
the modules, as in~\eqref{massive-modules1}--\eqref{massive-modules2})
\begin{multline}\label{mixed-2}
  \ldots
  \longrightarrow \mV_{\frac{\q}{p}(r+1) + 2\q m;-r-m p}\oplus
  \mU_{\frac{\q}{p}(r+1) - 2\q(m+1),0;(m+1)p - r}\oplus
  \mV_{\frac{\q}{p}(-r+1)-1-2\q m;m p} \longrightarrow{}\\
  \longrightarrow \mV_{-\frac{\q}{p}(r-1) + 2\q m;-m p}\oplus
  \mU_{-\frac{\q}{p}(r-1)-2\q m,0;m p}\oplus
  \mV_{\frac{\q}{p}(r+1)-1-2\q m;m p - r}
  \longrightarrow\\
  \ldots\\
  \longrightarrow\mV_{\frac{\q}{p}(r+1) + 2\q;-r-p}
  \oplus\mU_{\frac{\q}{p}(r+1)-4\q,0;2p - r}\oplus
  \mV_{\frac{\q}{p}(-r+1)-1-2\q;p}
  \longrightarrow{}\\
  \longrightarrow \mV_{-\frac{\q}{p}(r-1) + 2\q;-p}\oplus
  \mU_{-\frac{\q}{p}(r-1)-2\q,0;p}\oplus
  \mV_{\frac{\q}{p}(r+1)-1-2\q;p - r}
  \longrightarrow\\
  \longrightarrow
  \mV_{\frac{\q}{p}(r+1);-r}\oplus \mU_{\frac{\q}{p}(r+1) - 2\q,0;p -
    r}\oplus \mV_{-\frac{\q}{p}(r-1) - 1}\longrightarrow
  \mU_{-\frac{\q}{p}(r-1),0}
  \longrightarrow\mAs_{r,1,p,\q}\longrightarrow 0\,.
\end{multline}

Finally, every twisted topological Verma module is the quotient of a
massive Verma module over its submodule which is a twisted topological
Verma module.  The following Lemma, which can be shown in a
straightforward way, gives resolutions of twisted topological Verma
modules in terms of twisted massive Verma modules (from now on, we
restore~$t$ in the notation for the modules).
\begin{Lemma}\label{top-resolved}
  Let $\mV_{h,t;\theta}$ be a twisted topological Verma module.  There
  are the exact sequences
  \begin{gather}
    \ldots \longrightarrow
    \mU_{h + 3 - \frac{2}{t},h + 3 - \frac{2}{t},t;\theta+1}
    \longrightarrow
    \mU_{h + 2 - \frac{2}{t},h + 2 - \frac{2}{t},t;\theta+1}
    \longrightarrow
    \mU_{h + 1 - \frac{2}{t},h + 1 - \frac{2}{t},t;\theta+1}
    \longrightarrow
    \mV_{h,t;\theta}\longrightarrow 0\,,
    \label{top-resolved1}\displaybreak[3]\\
    0\longleftarrow\mV_{h,t;\theta}\longleftarrow
    \mU_{h,0,t;\theta}\longleftarrow
    \mU_{h-1,0,t;\theta}\longleftarrow
    \mU_{h-2,0,t;\theta}\longleftarrow\ldots\,.
    \label{top-resolved2}
  \end{gather}
\end{Lemma}
Using these in \eqref{mixed-2} for the {\Verma} and {\TVerma} modules,
respectively, we arrive at the resolution~\eqref{massive-resolution}.

\subsection{The massive-admissible
  representations\label{sec:massive-admissible}} Along with the
admissible $\N2$ representations defined above, one can consider a
more `exotic' case where an irreducible $\N2$ representation is
obtained by taking the quotient of a massive Verma module with respect
to only massive Verma submodules (that is, one starts with a massive
Verma module without charged singular vectors).
A characteristic feature of this situation is that all submodules can
be considered to have zero twist (because none of them have charged
singular vectors either).  The embedding diagram of the massive Verma
module in question is case III${}_+(0)$ of~\cite{[SSi]}. Then,
similarly to the standard admissible case, we can single out the
``massive-admissible'' representations as the quotients of those
massive Verma modules that are not embedded into another massive Verma
module.  These are the modules
$\mU_{h,\theell(r,s,h,\frac{p}{\q}),\frac{p}{\q}}$, where $h\in\oC$
and
\begin{equation}\label{theell}
  \theell(r,s,h,\tfrac{p}{\q})= -\tfrac {p'}{4p}
  \Bigl(\bigl(h\tfrac{p}{p'}-1\bigr)^2 -
  \bigl(r - \tfrac{p}{p'}(s-1)\bigr)^2\Bigr).
\end{equation}
The resolution reads as
\begin{multline}\label{massive-adm-resolution}
  \ldots \to \mN_+(m) \oplus \mN_-(m) \to
  \ldots\to\mN_+(2) \oplus \mN_-(2) \to \mN_+(1) \oplus
  \mN_-(1)\to{}\\
  {}\to\mU_{h,\theell(r,s,h,\frac{p}{\q}),\frac{p}{\q}}
  \to\mAm_{r,s,h,p,\q}\to0\,.
\end{multline}
where $\mN_{\pm}(m)=\mU_{h, -\frac{p'}{4p} \left((h\frac p{p'} -1)^2 -
    \Delta_{\pm}(m) ^2 \right) , \frac p{p'} }$ are the massive Verma
modules with
\begin{align*}
  \Delta_+ (2m)  &= r- \tfrac{p}{p'} (s-1) - 2pm\,,&\quad
  \Delta_- (2m) &= r- \tfrac{p}{p'} (s-1) + 2pm \,, \\
  \Delta_+ (2m-1) &= -r- \tfrac{p}{p'} (s-1) - 2p(m-1)\,,&
  \Delta_- (2m-1) &= -r- \tfrac{p}{p'} (s-1) + 2pm\,,
\end{align*}
where $m\in \oN$.  Since, as we have noted, all submodules in
$\mU_{h,\theell(r,s,h,\frac{p}{\q}),\frac{p}{\q}}$ are untwisted
massive Verma modules, the value of $h$ does not change along the
resolution (the entire resolution being thus, in a sense, a ``Virasoro
effect'').

\section{Characters of irreducible
  representations\label{sec:irreducible}} In this section, we use the
resolution constructed above to derive characters of the admissible
$\N2$ representations and to study their properties.

\subsection{The $\protect\tSL2$ side} We start with rederiving the
known results for the $\tSL2$ characters in order to show the
similarities (in fact, uniformity) with the $\N2$ ones and to find the
behaviour of characters under the spectral flow.
\subsubsection{Admissible representations} We  consider the
admissible representations as defined in~Sec.~{\it\ref{sec:BGG}}.  The
Sugawara dimension of the \hw{} state $|\jtop(r,s,\frac{p}{\q}),
\frac{p}{\q}\rangle_{s\ell(2)}$ then equals
\begin{equation}
  \tfrac{j(j+1)}{t}=
  \tfrac{(r^2-1)\q }{4p} - \tfrac{r}{2}(s-1) + \tfrac{(s-1)^2p}{4\q}\,.
\end{equation}
The character can be found from the BGG-resolution in a standard
manner.  It converges in the annulus $1<|z|<|q|^{-1}$, where it can be
summed to the character function
\begin{multline}\label{admissible-sl2}
  \charsl{\mAs}{r,s,p,\q}(z,q) =
  \frac{q^{\frac{(r^2-1)\q }{4p} - \frac{r}{2}(s-1) +
      \frac{(s-1)^2p}{4\q}
      + \frac{1}{8} - \frac{1}{4}p\q}}{
    \vartheta_{1,1}(z, q)}\,
  z^{\frac{r-1}{2} - (s-1)\frac{p}{2\q }}\times{}\\
  \times\left(
    \vartheta_{1,0}(z^p\,q^{-p\q  +  r \q  - (s-1) p}, q^{2p\q })
    {}-z^{-r}\,q^{r(s-1)}\,
    \vartheta_{1,0}(z^p\,q^{-p\q  -  r \q  - (s-1) p}, q^{2p\q })
  \right).
\end{multline}
\begin{Rem}
  Whenever $p=2r$, we use~\eqref{sl2trivial} to rewrite the above
  formula as
  \begin{equation}\label{p=2r-sl2}
    \charsl{\mAs}{r,s,2r,\q}(z,q) =
    q^{-\frac{\q}{8r} + \frac{(s-1)^2 r}{2\q }
      + \frac{1}{8} - \frac{r}{2}(s-1)}\,
    z^{\frac{r-1}{2} - \frac{r}{\q}(s-1) }\,
    \frac{\vt{1,1}{z^r\,q^{-r(s-1)}}{q^{r\q}}}{
      \vt{1,1}{z}{q}}\,.
  \end{equation}
\end{Rem}

A direct calculation (using the formulae of Appendix~\ref{app:theta})
shows that the admissible $\tSL2$ character functions are invariant
under the spectral flow with $\theta=2\q ${\rm:}
\begin{equation}
  \charsl{\mAs}{r,s,p,\q;2 \q}(z,q)=\charsl{\mAs}{r,s,p,\q}(z,q)\,,
  \qquad z\in\oC\,.
\end{equation}
Thus, explicitly,
\begin{equation}\label{section}
  \charsl{\mAs}{r,s,p,\q}(z\,q^{-2\q},q)=
  z^{p-2\q }\,q^{2{\q}^2 - p\q}
  \charsl{\mAs}{r,s,p,\q}(z,q)\,.
\end{equation}
In addition, whenever $p=2r$, we have
$\charsl{\mAs}{r,s,2r,\q;\q}(z,q)=(-1)^{p'-1}
\charsl{\mAs}{r,s,2r,\q}(z,q)$.

In view of this quasiperiodicity, the distinct character functions
are found by applying the spectral flow transform with
$0\leq\theta\leq2\q-1$ whenever $p\neq2r$, and with
$0\leq\theta\leq\q-1$ when $p=2r$. For such $\theta$, we find the
spectral-flow transformed character functions as
\begin{multline}\label{specadmissible-sl2}
  \charsl{\mAs}{r,s,p,\q;\theta}(z,q) =
  (-1)^\theta\,\frac{q^{\frac{ (r^2-1)\q }{4p} +
      \frac{p}{4\q }(s-1)(s+2\theta-1) - \frac{r}{2}(s+\theta-1)
      + \frac{1}{8} - \frac{1}{4}p\q + \half\theta^2}}{
    \vartheta_{1,1}(z, q)}\,
  z^{\frac{r-1}{2} - (s-1)\frac{p}{2\q } - \theta}\times{}\\
  \times
  \left(
    \vartheta_{1,0}(z^p\,q^{-p\q + r \q - (s+\theta-1) p}, q^{2p\q})
    -z^{-r}\,
    q^{r(s+\theta-1)}\,
    \vartheta_{1,0}(z^p\,q^{-p\q - r \q - (s+\theta-1) p}, q^{2p\q })
  \right).
\end{multline}

To summarise, while generically the spectral flow transform action on
twisted characters is
\begin{equation}\label{sl2-sf-trivial}
  \spfsl{\theta' }\charsl{\mAs}{r,s,p,p';\theta}(z,q)=
    \charsl{\mAs}{r,s,p,p';\theta+\theta'}(z,q)\,,
\end{equation}
we have the special cases described as follows.
\begin{Thm}\label{autadmrep:thm}  Character
  functions~\eqref{specadmissible-sl2} of the admissible $\tSL2$
  representations carry a representation of the group~$\Autsl$ given
  by~\eqref{sl2-sf-trivial} and
  \begin{align}
    \spfsl{2p'}\charsl{\mAs}{r,s,p,p';\theta}(z,q)&=
    \charsl{\mAs}{r,s,p,p';\theta}(z,q)\,,\label{work201}
    \displaybreak[3]\\
    \spfsl{p'}\charsl{\mAs}{r,s,p,p';\theta}(z,q)&=(-1)^{p'-1}
    \charsl{\mAs}{p-r,s,p,p';\theta}(z,q)\,,\label{work202}
    \displaybreak[3]\\
    \chevsl
    \charsl{\mAs}{ r,s,p,p';\theta } (z,q)&=(-1)^{p'-1}
    \charsl{\mAs}{r,p'-s+1,p,p';p'-\theta-1}(z,q)\,.
  \end{align}
\end{Thm}

\begin{Rem}\label{rem:puzzling}
  Equations~\eqref{work201} and~\eqref{work202} may seem puzzling if
  we recall that the twisted admissible representations certainly
  contain states in the bigradings\footnote{with respect to (level,
    charge), i.e., the eigenvalues of (the zero mode of) the Sugawara
    \emt{} and $J^0_0$, respectively.} in which, e.g., the untwisted
  representation contains no states.  However, in order to derive from
  a given character function $\chi(z,q)$ the occupation number
  $\chi_{m,n}$ in grade $(m,n)$, we have to expand it as
  $\chi(z,q)=\sum_{m,n}\chi_{m,n}z^m\,q^n$ in the annulus determined
  by the chosen representation. The expansion would depend on the
  annulus as soon as the character function has poles in~$z$.
\end{Rem}

As regards the poles of~$\charsl{\mAs}{r,s,p,\q}$, we have
\begin{Lemma}\label{admissible-poles} As a function of $z\in\oC$,
  the character function of the admissible $\tSL2$
  representation~$\mAs_{r,s,p,\q}$ has poles at the points $z=q^n$,
  $n\in \oZ\setminus(\q\oZ + s - 1)$.
\end{Lemma}
\begin{proof}
  The denominator of $\charsl{\mAs}{r,s,p,\q}(z,q)$ has simple zeros
  at $z=q^n$, $n \in \oZ$. We parametrise $n\mod\q$ as $n=\q \ell + s
  - 1 + a$ with $\ell\in\oZ$, $a=0,1,\ldots,\q-1$, and evaluate the
  bracket in~\eqref{admissible-sl2} at $z=q^n$:
  \begin{equation}
    \Bigl(~~\Bigr)=\sum_{m\in\oZ}
    q^{p\q m^2 - p\q \ell m - p a m - r\q m + \frac{r\q \ell}{2} +
      \frac{r a}{2}} -
    \sum_{m\in\oZ}
    q^{p\q m^2 - p\q \ell m - p a m + r\q m - \frac{r\q \ell}{2} -
      \frac{r a}{2}}.
  \end{equation}
  Replacing $m\mapsto \ell-m$ in the second sum, we rewrite this as
  \begin{equation}
    \sum_{m\in\oZ}
    q^{p\q m^2 - p\q \ell m - r\q m + \frac{r\q \ell}{2}}
    \left(
      q^{\frac{r a}{2} - p a m} - q^{-\frac{r a}{2} + p a m - \ell p a}
    \right),
  \end{equation}
  which vanishes whenever $a=0$.
\end{proof}

\subsubsection{Unitary representations} A module $\mL$ is called
unitary if the Shapovalov form~$\shpv\cdot\cdot$ is positive definite.
We now consider those admissible representations that are unitary,
namely the ones with $s=\q=1$; these are the {\it integrable\/}
representations (thus, our usage of the term ``unitary'' is limited to
the unitary admissible representations). The characters of these
representation are therefore given by
\begin{equation}\label{integrable-sl2}
  \charsl{\mL}{r,p}(z,q) =
  \frac{q^{\frac{r^2-1}{4p} + \frac{1}{8} - \frac{p}{4}}}{
    \vartheta_{1,1}(z, q)}\,z^{\frac{r-1}{2}}\,
  \left(
    \vartheta_{1,0}(z^p\,q^{-p + r}, q^{2p}) -
    z^{-r}
    \vartheta_{1,0}(z^p\,q^{-p - r}, q^{2p})
  \right)\,.
\end{equation}
\begin{Lemma}
  The character function of the unitary $\tSL2$ representation is
  holomorphic in~$z\in\oC$.
\end{Lemma}

The modular transformation properties of these characters are
well-known, and we do not repeat them here. Instead, we point out how
the characters behave under the spectral flow and the involution:
\begin{Lemma} The character functions of the unitary $\tSL2$ modules
  carry the following representation of the group~$\Autsl$:
\begin{alignat}{1}
    \spfsl{1}\charsl{\mL}{r,p}(z,q)&=\charsl{\mL}{p-r,p}(z,q)\,,\\
    \chevsl\charsl{\mL}{r,p}(z,q)&=\charsl{\mL}{r,p}(z,q)\,.
  \end{alignat}
  In particular, in the case where $p=2r$, we have
  $\spfsl{1}\charsl{\mL}{r,p}(z,q)=\charsl{\mL}{r,p}(z,q)$.
\end{Lemma}
\noindent
We thus see that
$\spfsl{2}\charsl{\mL}{r,p}(z,q)=\charsl{\mL}{r,p}(z,q)$, i.e., all
the unitary characters are invariant under the spectral flow with even
transformation parameters (not surprisingly, since $(2\oZ)\simtimesl
\oZ_2\subset\Autsl$ is the affine Weyl group).
\begin{Rem}
  Unlike the case with admissible representations, it is not only the
  characters but actually the unitary {\it representations\/}
  themselves that are mapped into each other by the group~$\Autsl$
  (this agrees with the fact that the unitary characters are
  holomorphic, cf.~{\it\ref{rem:puzzling}\/}).  It is easy to see how
  the unitary representations are mapped by the spectral flow at the
  level of their {\it extremal diagrams\/}~\cite{[FST]}. \ Such
  representations are the quotients of Verma modules with singular
  vectors of the form $(J^-_0)^r$ and $(J^+_{-1})^{p-r}$ (which are
  understood to act on the \hw{} vector).  In the generic case where
  $r\neq1$ or $p-1$, the extremal diagram of the unitary
  representation $\mL_{r,p}$ looks like (in the conventions
  of~\cite{[FST]})
  \begin{equation}\label{diag:sl2-unitary}
    \begin{picture}(400,115)
      \put(-20,0){
      \put(250,100){\vector(-1,0){260}}
      \put(250,100){\vector(2,-1){180}}
      {\thicklines
        \put(325.0,40.0){\line(1,-2){16}}
        \put(325.2,39.8){\line(1,-2){16}}
        \put(280.2,85.0){\line(1,-1){45}}
        \put(280,84.8){\line(1,-1){45}}
        \put(250.2,100){\line(2,-1){30}}
        \put(250,99.8){\line(2,-1){30}}
        \put(250,99.7){\line(-1,0){60}}
        \put(250,100.1){\line(-1,0){60}}
        \put(190.2,100){\line(-2,-1){30}}
        \put(190,99.8){\line(-2,-1){30}}
        \put(160.2,85.0){\line(-1,-1){45}}
        \put(160,84.8){\line(-1,-1){45}}
        \put(115.3,40){\line(-1,-2){16}}
        \put(115.1,39.8){\line(-1,-2){16}}
        }
      \put(215,105.7){$\scriptstyle\bigl(J^-_{0}\bigr)^{r-1}$}
      \put(267,98.7){$\scriptstyle\bigl(J^-_{1}\bigr)^{p-r-1}$}
      \put(180,83.7){$\scriptstyle\bigl(J^-_{-1}\bigr)^{p-r-1}$}
      \put(261,55.7){$\scriptstyle\bigl(J^-_{2}\bigr)^{r-1}$}
      \put(145,55.7){$\scriptstyle\bigl(J^-_{-2}\bigr)^{r-1}$}
      \put(284,15.7){$\scriptstyle\bigl(J^-_{3}\bigr)^{p-r-1}$}
      \put(115,15.7){$\scriptstyle\bigl(J^-_{-3}\bigr)^{p-r-1}$}
      }
    \end{picture}
  \end{equation}
  The sections between the cusps are such that their horizontal
  projections (onto the charge axis) have lengths \ldots, $p-r-1$,
  $r-1$, $p-r-1$, $r-1$, \ldots.  Thus, marking each section with the
  action of a $J^-_n$ operator, the extremal diagram of $\mL_{r,p}$
  can be described as
  \begin{equation}
    \ldots\ \bigl(J^-_{i-3}\bigr)^{p-r-1}\bigl(J^-_{i-2}\bigr)^{r-1}
    \bigl(J^-_{i-1}\bigr)^{p-r-1}\bigl(J^-_{i}\bigr)^{r-1}
    \bigl(J^-_{i+1}\bigr)^{p-r-1}
    \ldots\ \longleftrightarrow\ \mL_{r,p}\,.
  \end{equation}
  After the spectral flow transformation~$\spfsl{1}$, this takes the
  form
  \begin{equation}
    \kern5pt\ldots\ \bigl(J^-_{i-3}\bigr)^{r-1}\bigl(J^-_{i-2}\bigr)^{p-r-1}
    \bigl(J^-_{i-1}\bigr)^{r-1}\bigl(J^-_{i}\bigr)^{p-r-1}\bigl(J^-_{i+1}\bigr)^{r-1}
    \ldots\ \longleftrightarrow\ \mL_{p-r,p}\,,
  \end{equation}
  in agreement with $\spfsl{1}\mL_{r,p}=\mL_{p-r,p}$.
\end{Rem}

\subsection{The $\N2$ side} We now derive the $\N2$ characters using
the BGG resolution constructed in Sec.~{\it\ref{sec:BGG}} and also the
massive resolutions from Sec.~{\it\ref{sec:massive}}.

\subsubsection{Admissible representations} Consider the admissible
$\N2$ representations as defined in~Sec.~{\it\ref{sec:resolutions}}.
We can start with the untwisted topological Verma modules and then
take the twist.  Let, thus, the admissible representation be
$\mAn_{r,s,p,\q}= \mV_{\htop(r,s,\frac{p}{\q}),\frac{p}{\q}}/{(\rm
  maximal\ submodule)}$.

The alternating sum of characters over the resolution
from Theorem~\ref{thm:resolution-N}
reads as
\begin{equation}
  \charn{\mAn}{r,s,p,\q }(z,q) =
  \sum_{m=-\infty}^\infty
  \charn{\mV}{-\frac{\q }{p}(r-1) + s-1 - 2\q  m,\frac{p}{\q };m p}(z,q) -
  \sum_{m=-\infty}^\infty
  \charn{\mV}{\frac{\q }{p}(r+1) + s-1 - 2\q  m,\frac{p}{\q };m p-r}(z,q)\,.
\end{equation}
As we have seen in Sec.~{\it\ref{BGG-N2}}, the twists
$\charn{\mV}{\ldots;m p}(z,q)$ and $\charn{\mV}{\ldots;m p-r}(z,q)$
follow from the fact that the corresponding submodules are generated
from twisted topological \hw{} states.  Explicitly substituting the
topological Verma module characters from~\eqref{topchar} leads to the
following Theorem.\footnote{See also the evaluation of the $s=1$
  characters in~\cite{[D-emb]}, where an exact sequence equivalent
  to~\eqref{mixed-2} appears to have implicitly been used, however
  with the modules being untwisted and generated from top-level
  singular vectors, which has required the unnecessary verification
  that the superfluous ``subsingular vectors'' are absent.}
\begin{Thm}
  The admissible $\N2$ characters read as
  \begin{equation}\label{admissible-N2}
    \charn{\mAn}{r,s,p,\q }(z,q)=\varphi_{r,s,p,\q}(z,q)\;
    \charn{\mU}{-\frac{\q}{p}(r-1)+s-1,0,\frac{p}{\q}}(z,q)\,,
  \end{equation}
  where
  \begin{equation}\label{massive-factor}
    \charn{\mU}{-\frac{\q}{p}(r-1)+s-1,0,\frac{p}{\q}}(z,q)=
    z^{-\frac{\q}{p}(r-1)+s-1}\,
    \frac{\vartheta_{1,0}(z,q)}{
      \eta(q)^3}\,,
  \end{equation}
  and
  \begin{equation}\label{varphi}
    \varphi_{r,s,p,\q}(z,q)=
    \sum_{m\in\oZ}
    q^{p\q m^2 - mp(s-1)}\left(
      \frac{q^{m\q r}}{1+z^{-1}\,q^{mp}} -
      q^{r(s-1)}\,\frac{q^{-m\q r}}{1+z^{-1}\,q^{mp-r}}
    \right).
  \end{equation}
\end{Thm}
\noindent
(We have separated into~\eqref{massive-factor} a factor equal to the
massive Verma module character).  The following statement is obvious:
\begin{Lemma}
  The admissible $\N2$ character functions are holomorphic
  in~$z\in\oC$.
\end{Lemma}

Next, we study the behaviour of $\charn{\mAn}{r,s,p,\q }$ under the
spectral flow. The transformation law~\eqref{spectral-n2-general}
induces a transformation of $\varphi_{r,s,p,\q}$, which we now
evaluate.  As follows by a direct calculation, the following
identities describe the behaviour of $\varphi_{r,s,p,\q}(z,q)$ under
the spectral flow transform: for $n\geq1$, we have
\begin{multline}\label{phi-identity}
  z^{2\q n}\,q^{\q n(r-p n) - p n(s-1)}
  \varphi_{r,s,p,\q}(z\,q^{-p n},q) - \varphi_{r,s,p,\q}(z,q)={}\\
    {}=\sum_{a=0}^{2\q n-1}
    (-1)^{a+1}\,z^{a+1}\,q^{-p\q/4}
    \Bigl(\vt{1,0}{q^{-p\q + p(s+a) - \q r}}{q^{2p\q}}
    -q^{r(s+a)}\,\vt{1,0}{q^{-p\q + p(s+a) + \q r}}{q^{2p\q}}\Bigr)\,.
\end{multline}
and for $n\leq-1$, similarly,
\begin{multline}\label{phi-identity2-}
  z^{2\q n}\,q^{\q n(r-p n) - p n(s-1)}
  \varphi_{r,s,p,\q}(z\,q^{-p n},q) - \varphi_{r,s,p,\q}(z,q)={}\\
  =\sum_{a=2\q n}^{-1}
  (-1)^{a}\,z^{a+1}\,
  q^{-p\q/4}
  \Bigl(\vt{1,0}{q^{-p\q + p(s+a) - \q r}}{q^{2p\q}}
  -q^{r(s+a)}\,\vt{1,0}{q^{-p\q + p(s+a) + \q r}}{q^{2p\q}}\Bigr)\,.
\end{multline}
In the same way as in the proof of Lemma~\ref{admissible-poles}, we
can see that those terms on the right-hand side
of~\eqref{phi-identity} where $a=\ell\q - s$, $1\leq\ell\leq2n$,
actually vanish.  In general, $\varphi_{r,s,p,\q}(z,q)$ is not
quasiperiodic in $z$ and, thus, cannot be algebraically expressed
through theta functions. An exception is provided by the unitary
representations considered below.
\begin{Rem}
  It is amusing to observe that the inhomogeneous terms in the
  transformation law of $\varphi_{r,s,p,\q}(z,q)$ under the spectral
  flow are
  the {\it residues\/} of the admissible $\tSL2$ characters; while
  presently this comes as an ``experimental fact'', we see in the next
  section why it must be so.
\end{Rem}
\begin{Rem}\label{rem:annuli}
  When $|q|^{-r+pn}<|z|<|q|^{-p+pn}$, Eq.~\eqref{varphi} can
  also be written as
  \begin{equation}\label{varphi3}
    \varphi_{r,s,p,\q}(z,q) = \Bigl(\sum_{\substack{m\geq n\\ k\geq0}}
    - \sum_{\substack{m\leq n-1\\ k\leq-1}}\Bigr)
    q^{p\q m^2 - m p(s-1) + m p k}(-1)^k\,z^{-k}\,\left(
      q^{m \q r} - q^{r(s-1) - m \q r - r k}
    \right).
  \end{equation}
  Similarly, when $|q|^{pn}<|z|<|q|^{-r+pn}$, we can expand as
  \begin{multline}\label{varphi3+1}
    \varphi_{r,s,p,\q}(z,q) = \Bigl(\sum_{\substack{m \geq n+1\\
        k\geq0}} - \sum_{\substack{m\leq n-1\\ k\leq-1}}\Bigr) q^{p\q
      m^2 - m p(s-1) + m p k}(-1)^k\,z^{-k}\,\left( q^{m \q r} -
      q^{r(s-1) - m \q r - r k}
    \right) + {}\\
    {} + q^{p\q n^2 - n p(s-1)} \Bigl( q^{n \q r}
    \sum_{\substack{k\geq0}} (-1)^k\,z^{-k}\,q^{n p k} + q^{r(s-1) - n
      \q r} \sum_{\substack{k\leq-1}} (-1)^k\,z^{-k}\,q^{n p k - r k}
    \Bigr).
  \end{multline}
  Using this in~\eqref{admissible-N2}, we recall the factor given by
  the character of a massive $\N2$ Verma module, which suggests that
  the character of an admissible representation is given by an
  infinite sum of massive Verma module characters. This is not
  accidental in view of the following result (even though it is by
  itself limited to the case of $s=1$).
\end{Rem}
\begin{Lemma}
  The character formula for the admissible $\N2$ representation
  $\mAs_{r,1,p,\q}$ following from the massive
  resolution~\eqref{massive-resolution} reads as~\eqref{admissible-N2}
  {\rm(}with $s=1${\rm)}, where
  \begin{multline}\label{char-new}
    \varphi_{r,1,p,\q}(z,q)={}\\
    \begin{split}
      {}=&\sum_{m\geq0}\biggl( \sum\limits_{j=1}^m
      z^{1+2(m-j)}q^{j[p\q j + p + 2p(m-j)-\q r]} +
      \sum\limits_{j=0}^{m} z^{2(m-j)}q^{(p j + r)[\q j + 2(m-j)]}
      \displaybreak[3]\\
      &\phantom{\sum_{m\geq0}\biggl(} {}+\sum\limits_{j=1}^m
      z^{-1-2(m-j)}q^{(pj-r)[\q j + 1 + 2(m-j)]}
      +\sum\limits_{j=0}^{m-1} z^{-2(m-j)}q^{j[p\q j + \q r +2p(m-j)]}
      \biggr)\\
      &{}-\sum_{m\geq0}\biggl( \sum\limits_{j=0}^m
      z^{1+2(m-j)}q^{(pj+r)[\q j + 1 + 2(m-j)]} +
      \sum\limits_{j=1}^{m+1} z^{2(1+m-j)}q^{j[p\q j + 2p(1+m-j)-\q r]}
      \\
      &\phantom{{}-\sum_{m\geq1}\biggl(} {}+\sum\limits_{j=0}^m
      z^{-1-2(m-j)}q^{j[p\q j + 2p(m-j) + p + \q r]}
      +\sum\limits_{j=1}^{m} z^{-2(1+m-j)}q^{(pj-r)[\q j + 2(1 + m
        -j)]} \biggr).
    \end{split}
  \end{multline}
  This converges for $1<|z|<|q|^{-r}$ and is then equal
  to~\eqref{varphi3+1}.
\end{Lemma}
\begin{proof} The character formula is straightforward to derive by
  inserting the twisted massive Verma module
  characters~\eqref{massive-character} into the massive
  resolution~\eqref{massive-resolution}--\eqref{massive-modules2}. \ 
  The organization of~\eqref{char-new} is in accordance with the
  3,\,5,\,7,\ldots structure of the resolution. For $1<|z|<|q|^{-r}$,
  the sum in~\eqref{char-new} can be brought to the
  form~\eqref{varphi3+1} by changing the order of summations in each
  term and then redefining the summation indices.
\end{proof}

\subsubsection{A series of unitary representations} A module $\mK$ is
called unitary if the form~\eqref{Shap-N2} is positive definite.
Taking the untwisted admissible representations with $s=\q=1$
\cite{[BFK]}, we obtain $p-1$ from the total of $\half p(p-1)$ unitary
representations (all of which are considered in
Sec.~{\it\ref{subsubsec:All-unitary}\/}).  Therefore, the characters
of the unitary representations $\mK_{r,p}\equiv\mAn_{r,1,p,1}$ are
given by
\begin{equation}\label{N2-unitary1}
  \charn{\mK}{r,p}(z,q)=
  \frac{\vartheta_{1,0}(z,q)}{
    \eta(q)^3}\,z^{-\frac{r-1}{p}
    }
  \sum_{m\in\oZ}
  q^{p m^2}\left(
    \frac{q^{m r}}{1+z^{-1}\,q^{m p}} -
    \frac{q^{m r}}{1 + z^{-1}\,q^{-m p-r}}
  \right)\,,\quad 1\leq r\leq p-1\,.
\end{equation}

Obviously, the character functions of unitary $\N2$ representations
are holomorphic in~$z\in\oC$.  Moreover,
\begin{Lemma} The characters of unitary $\N2$ representations are
  invariant under the spectral flow transform with $\theta=p$\,{\rm:}
  \begin{equation}\label{work-lemma-1}
    \charn{\mK}{r,p;p}(z,q) = \charn{\mK}{r,p}(z,q)\,.
  \end{equation}
  In addition, whenever $p=2r$, the character is invariant under the
  spectral flow with $\theta=r$:
  \begin{equation}\label{work-lemma-2}
    \charn{\mK}{r,2r;r}(z,q) = \charn{\mK}{r,2r}(z,q)\,.
  \end{equation}
\end{Lemma}
\begin{proof}
  We recall that the right-hand side of~\eqref{phi-identity} (where
  now $n=1$) vanishes when $a=\q - s$ and $a=2\q - s$. In the unitary
  case, where $\q=s=1$, this means that these terms vanish altogether,
  and therefore,
  \begin{equation}\label{work505}
    z^{2}\,q^{r-p}\,\varphi_{r,1,p,1}(z\,q^{-p},q) =
    \varphi_{r,1,p,1}(z,q)\,,
  \end{equation}
  which shows \eqref{work-lemma-1}; the second assertion is proved
  similarly.
\end{proof}
\noindent
Explicitly,
\begin{equation}\label{periodicity-N2-full}
  \charn{\mK}{r,p}(z\,q^{-p n},q) = z^{n(p-2)}\,
  q^{-\frac{1}{2}(p-2)n(pn-1)}\,
  \charn{\mK}{r,p}(z,q)\,.
\end{equation}

Thus, $\charn{\mK}{r,p}(z,q)$ are quasiperiodic holomorphic functions
of $z\in\oC$, and therefore, they can be expressed through the theta
functions.  This allows us to arrive at the following
statement~\cite{[M],[KW]}:
\begin{Thm} The characters of the unitary $\N2$ representations
  $\mK_{r,p}\equiv\mAn_{r,1,p,1}$ are given by
  \begin{equation}\label{N2-unitary}
    \charn{\mK}{r,p}(z,q)=
    -z^{-\frac{r-1}{p}-1}\,                   
    \frac{\eta(q^p)^3}{\eta(q)^3}\,
    \frac{\vt{1,0}{z}{q}\,\vt{1,1}{q^r}{q^p}}{
      \vt{1,0}{z}{q^p}\,\vt{1,0}{z\,q^r}{q^p}}\,.
  \end{equation}
\end{Thm}
\begin{Rem}
  In the special case where $p=2r$, the last formula rewrites as
  \begin{equation}\label{p=2r-N2}
    \charn{\mK}{r,2r}(z,q)=
    z^{-\frac{r-1}{2r}}\,                     
   \frac{\eta(q^r)^3}{\eta(q)^3}\,
    \frac{\vt{1,0}{z}{q}}{
      \vt{1,0}{z}{q^r}}\,.
  \end{equation}
\end{Rem}
\begin{proof}
  We see from~\eqref{work505} that
  \begin{align}\label{N2-unitary0}
    \varphi_{r,1,p,1}(z,q)&=\frac{f_{r,p}(q)}{
      z\vt{1,0}{z}{q^p}\,\vt{1,0}{z\,q^r}{q^p}}\,.
  \end{align}
  Indeed, the right-hand side of \eqref{N2-unitary0} considered as a
  function of $z$ has all the poles of~\eqref{varphi} and, for
  $s=\q=1$, possesses the same periodicity properties. To find
  $f_{r,p}(q)$, we evaluate the residues of both sides of
  \eqref{N2-unitary0} at $z=-q^{p n}$, $n\in\oZ$
  using~\eqref{theta-prime}:
  \begin{gather*}
    \res_{z=-q^{p n}} \sum_{m\in\oZ}q^{p m^2} \left( \frac{q^{m r}}{1
        + z^{-1}\,q^{m p}} - \frac{q^{m r}}{1 + z^{-1}\,q^{-m p - r}}
    \right) =
    -q^{p n^2 + p n  + n r}\,,\\
    \res_{z=-q^{p n}} \frac{f_{r,p}(q)}{z\,\vt{1,0}{z}{q^p}\,
      \vt{1,0}{z\,q^r}{q^p}} =
    f_{r,p}(q)\, \frac{q^{p n^2 + p n + r       
        n}}{\eta(q^p)^3 \vt{1,1}{q^r}{q^p}}\,.
  \end{gather*}
  Therefore, $f_{r,p}(q) = -\eta(q^p)^3\, \vt{1,1}{q^r}{q^p}$.
\end{proof}
For the modular properties of these characters, see~\cite{[RY],[KW]}.

\subsubsection{All unitary
  representations\label{subsubsec:All-unitary}} We have seen that the
unitary $\N2$ representations can be obtained by taking quotients of
twisted topological Verma modules~$\mV_{h,p;\theta}$ with
$h=\frac{1-r}{p}$, where 
$1\leq r\leq p-1$, $p-2\in\oN$, over two topological singular vectors.
The unitary modules can be also obtained by taking the quotients of
the massive Verma modules~$\mU_{h,\ell,p}$ with $h=\frac{1+n-m}{p}$
and $\ell=\frac{mn}{p}$, where $1\leq n+m\leq p-1$, $1\leq m$, and
$0\leq n$, over two charged (see~\eqref{ECh}) and one massive singular
vectors.  The $r$ and~$\theta$ parameters are then recovered as
$r=n+m$ and $\theta=n$.  The \hw{} vector of the quotient module lies
on the extremal diagram of the massive Verma module and is a {\it
  twisted\/} topological \hw{} vector.  Different twists give rise to
generically non-isomorphic representations, however there are the
isomorphisms
\begin{equation}\label{n2-isomorphisms}
  \mK_{r,p;\theta+p}\approx\mK_{r,p;\theta}\quad
  \text{and}\quad
  \mK_{r,p;\theta+r}\approx\mK_{p-r,p;\theta}\,.
\end{equation}
\vbox{\vskip2.2\baselineskip
{\hfill\unitlength=1pt
  \begin{picture}(200,160)\label{boxes}
  \multiput(0,0)(20,0){5}{
    \line(0,1){140}
    }
  \put(90,10){\Large\dots}
  \multiput(120,0)(20,0){4}{
    \line(0,1){140}
    }
  \multiput(0,0)(0,20){4}{
    \line(1,0){180}
    }
  \put(100,80){\Large\vdots}
  \multiput(0,100)(0,20){3}{
    \line(1,0){180}
    }
  {\thicklines
    \put(3.5,0){\vector(1,0){200}}
    \put(3.5,0){\vector(0,1){155}}
    }
  \put(196,5){$\theta$}
  \put(8,150){$r$}
  \put(28,10){${}_{[11]}$}\put(8,130){${}_{[11]}$}
  \put(48,29){${}_{[22]}$}\put(8,110){${}_{[22]}$}
  \put(48,10){${}_{[21]}$}\put(28,130){${}_{[21]}$}
  \put(68,10){${}_{[31]}$}\put(48,130){${}_{[31]}$}
  \put(68,30){${}_{[32]}$}\put(28,110){${}_{[32]}$}
  \put(165,13){${}_{\substack{[p-1\\,1]}}$}
    \put(145,133){${}_{\substack{[p-1\\,1]}}$}
  \put(8,10){${}_{[01]}$}\put(168,130){${}_{[01]}$}
\end{picture}
}

\vskip-11.8\baselineskip
\noindent
\vtop{\parshape 13
0pt \textwidth
0pt \textwidth
0pt .55\textwidth
0pt .55\textwidth
0pt .55\textwidth
0pt .55\textwidth
0pt .55\textwidth
0pt .55\textwidth
0pt .55\textwidth
0pt .55\textwidth
0pt .55\textwidth
0pt .55\textwidth
0pt \textwidth
\noindent
The first of these isomorphisms means periodicity with respect to the
spectral flow with period~$p$ (therefore, we can take
$\theta\in\oZ_p$), while the second one shows that in the table
consisting of $p(p-1)$ boxes that correspond to $0\leq\theta\leq p-1$
and $1\leq r\leq p-1$, there are $\half p(p-1)$ non-isomorphic unitary
representations.  Some of the isomorphisms described
by~\eqref{n2-isomorphisms} are shown in the diagram (the boxes with
the same values of $[\theta r]$). Whenever $p$ is even, the row
labelled by $r=p/2$ is obtained by repeating twice the first $p/2$
representations, i.e., it reads as
$([0,\frac{p}{2}],\ldots,[\frac{p}{2}-1,\frac{p}{2}],\linebreak[3]
[0,\frac{p}{2}],\ldots,[\frac{p}{2}-1,\frac{p}{2}])$.  If the $r$
and~$\theta$ parameters are expressed through $n$ and $m$ as above, it
follows that $0\leq\theta\leq r-1$, thereby ensuring that each unitary
representation is counted precisely once.}
}

\vskip.8\baselineskip

We now outline an alternative description of the unitary
modules~\cite{[FST-semi]} (which is a counterpart of the $\tSL2$
unitary module construction~\cite{[FSt]}).  The operator
$S(z)=\d^{p-2}\cG(z)\dots\d\cG(z)\cG(z)$ is a {\it null field\/}
(because $S(0)\ket{0}$ is a singular vector in the vacuum
representation) and is therefore quotiened away in any unitary
representation.  This gives the relations
\begin{equation}\label{Q-relation}
  \sum_{\substack{i_0<\dots<i_{p-2}\\ i_0+\dots+i_{p-2}=a}}
  \Bigl(\prod_{m<n}(i_m-i_n)\Bigr)
  \cG_{i_0}\dots\cG_{i_{p-2}}=0\,,\qquad a\in\oZ\,
\end{equation}
that hold in every unitary module.  These allow us to describe the
structure of the extremal diagram of unitary modules.  Every product
\begin{equation}\label{G-product}
  \cG_{a_1}\ldots\cG_{a_N}\,,\qquad a_1<\ldots<a_N
\end{equation}
(where the ordering can always be achieved at the expense of an
overall sign) can be encoded by saying which of the numbers from
$\{a_1,a_1+1,\ldots,a_N-1,a_N\}$ are {\it occupied}, i.e., are such
that the corresponding mode of $\cG$ is present in~\eqref{G-product}.
Now, when constructing extremal states \ 
$\dots\cG_{\theta-2}\cG_{\theta-1}$ (which is assumed to act on state
satisfying the $\theta$-twisted topological \hw{} conditions), every
such product vanishes as soon as there are $p-1$ consecutive numbers
that are occupied in~$\{a_1,a_1+1,\ldots,a_N-1,a_N\}$.

The unitary extremal diagram can be described by periodic sequences
with period $p$ consisting of $\times$ or~$\circ$ (occupied/unoccupied
positions) and exactly $p-2$ occupied numbers (the crosses) per
period:
\begin{equation}
  \begin{array}{l}
    \ldots\,
    \overbrace{\times\dots\times\ldots\times}^{r-1}\circ
    \overbrace{\times\dots\times}^{p-r-1}\circ\,\ldots\\[-2pt]
    \phantom{\ldots\,\times\dots{}}\,{}^\wedge
  \end{array}
\end{equation}
One of the positions is marked as the basepoint.  Two such periodic
infinite sequences are identified whenever they coincide (including
their basepoints) after a translation by $ip$, $i\in\oZ$.

With the modes $\cQ_{-n}$ on the $\cQ$-side of the extremal diagram
formally replaced with $\cG_{n}$, such sequences are in a $1:1$
correspondence with extremal diagrams of the unitary modules; the
unoccupied positions correspond to cusps in the extremal diagram, and
the basepoint is placed to grade 0 with respect to $\cH_0$:
\begin{equation}\label{extremal-unitary}
  \unitlength=.8pt
  \begin{picture}(500,135)
    \put(60,0){
      {\linethickness{.4pt}
        \qbezier[1000](80,0)(200,280)(320,0)
        }
      \linethickness{.9pt}
      \qbezier[200](120,30)(125,50)(140,90)
      \qbezier[200](140,90)(155,120)(179,136)
      \qbezier[200](179,136)(210,148)(240,124)
      \qbezier[200](240,124)(256,110)(270,80)
      \qbezier[200](270,80)(280,50)(290,10)
      }
  \end{picture}
\end{equation}
The outer parabola is the extremal diagram of a massive Verma module,
which also can be represented as a {\it dense fill\/}
\begin{equation*}
  \ldots\,\times\,\times\,\times\,\ldots\,.
\end{equation*}
Moreover, the basis for {\it all\/} states in the unitary module is
given by a semi-infinite construction where a finite number of
length-$p$ fragments contain {\it not more\/} than $p-2$ occupied
positions each, while the infinite tail
$\cG_{a_1}\ldots\cG_{a_{i-1}}\cG_{a_i}\ldots$ with
$a_1<\ldots<a_{i-1}<a_i<\ldots$, is such that the associated sequence
of the occupied positions is periodic with period $p$ and has exactly
$p-2$ occupied positions per period~\cite{[FST-semi]}.

An infinite $p$-periodic sequence (with a basepoint) consisting of
$\times$ and $\circ$ with $p-2$ $\times$'s per period can be
characterised by its length-$p$ section starting at the basepoint,
\begin{equation}\label{abc}
  \begin{array}{l}
    \overbrace{\times\times\ldots\times}^a
    \circ
    \overbrace{\times\ldots\times}^b
    \circ
    \overbrace{\times\ldots\times}^c\\[-2pt]
    \,{}^\wedge
  \end{array}
\end{equation}
where $0\leq a\leq p-2$, $0\leq c\leq p-2$, $0\leq a+c\leq p-2$ (and
$a+b+c=p-2$). This corresponds to the unitary representation
$\mK_{r,p;\theta}$ with $r=a+c+1$ and $\theta=a$.  We note again that
this implies $0\leq\theta\leq r-1$ with $1\leq r\leq p-1$, and,
therefore, each unitary representation is labelled once.

For example, in the case of $p=4$ the possible extremal diagrams are
\begin{alignat*}{3}
  &\ldots\,\circ\,\circ\,\times\,\times\,\ldots\qquad
  &
  &\ldots\,\circ\,\times\,\circ\,\times\,\ldots\qquad
  &
  &\ldots\,\circ\,\times\,\times\,\circ\,\ldots\\[-6pt]
  &\kern19pt{}^\wedge
  &
  &\kern19pt{}^\wedge
  &
  &\kern20pt{}^\wedge\\
  &\ldots\,\times\,\times\,\circ\,\circ\,\ldots
  &                                     
  &\ldots\,\times\,\circ\,\times\,\circ\,\ldots
  &                                     
  &\ldots\,\times\,\circ\,\circ\,\times\,\ldots\\[-6pt]
  &\kern20pt{}^\wedge
  &
  &\kern20pt{}^\wedge
  &
  &\kern20pt{}^\wedge
\end{alignat*}

A unitary module is characterised by its extremal diagram.
\begin{Lemma}
  There exist $\half p(p-1)$ different $p$-periodic infinite sequences
  $(\lambda_i)_{i\in\oZ}$, $\lambda_i\in\{\times,\circ\}$ with a
  basepoint and with $p-2$ crosses~$\times$ per period.  Thus, there
  exist $\half p(p-1)$ non-isomorphic unitary $\N2$ representations
  with the central charge $\ctop=3(1-\frac{2}{p})$, $p\in\oN$,
  $p\geq2$.
\end{Lemma}

The characters of all the $p(p-1)/2$ inequivalent unitary
representations are found by taking the twists of the above
characters~\eqref{N2-unitary}.  This is done in accordance with
formula~\eqref{spectral-n2-general} for the spectral flow transform of
the $\N2$ characters.  While generically one has
\begin{equation}\label{spectral-u-gen}
  \spfn{\theta'}\charn{\mK}{r,p;\theta}(z,q)=
  \charn{\mK}{r,p;\theta+\theta'}(z,q)\,,
\end{equation}
we see that there are the following special cases of the spectral flow
transform, in accordance with~\eqref{n2-isomorphisms}:
\begin{align}
  \spfn{p}\charn{\mK}{r,p;\theta}(z,q)&
  =\charn{\mK}{r,p;\theta}(z,q)\,,\label{N2-spec-1}\\
  \spfn{r}\charn{\mK}{r,p;\theta}(z,q)&
  =\charn{\mK}{p-r,p;\theta}(z,q)\,.
  \label{N2-spec-2}
\end{align}
When, in particular, $p=2r$, we have
$\spfn{r}\charn{\mK}{r,2r;\theta}(z,q)=\charn{\mK}{r,2r;\theta}(z,q)$.

\begin{Rem}
  Note that {\it if we identify\/} the representations that differ by
  the spectral flow, we will be left with only $[p/2]$ classes of
  unitary $\N2$ representations, which is the same as the number of
  classes of unitary $\tSL2$ representations modulo the spectral
  flow~\cite{[FST]}.
\end{Rem}

\subsubsection{More on the admissible $\N2$ characters} Unlike
the unitary representations, the characters of general admissible
representations, which are {\it not\/} quasi-periodic, cannot be
expressed algebraically in terms of a finite number of theta
functions.  In Sec.~{\it\ref{sec:admissible-branching}}, we consider
an integral representation for the admissible $\N2$ characters, while
now we briefly comment on how one can arrive at equations for these
characters.

Observe first of all that although setting $\q=1$ implies that $s=1$
in the case of admissible representations, we can formally define
$\varphi_{r,s,p,1}(z,q)$ by the same formula~\eqref{varphi} even for
$s>1$. Since $\varphi_{r,s,p,1}(z,q)$ is still quasi-periodic under
$z\mapsto z\,q^p$, we can proceed similarly to the above to derive
\begin{equation*}
  \varphi_{r,s,p,1}(z,q)=
  -\frac{\eta(q^p)^3\,                           
    \vt{1,1}{q^r}{q^p}}{
    z^s\;\vt{1,0}{z}{q^p}\,\vt{1,0}{z\,q^r}{q^p}}\,.
\end{equation*}
Now, let $e_{\q}=e^{{2\pi i}/{\q}}$ be the primitive $\q$th root of
unity.  We then see that
\begin{equation*}
  \sum_{a=0}^{\q-1}\,\varphi_{r,s,p,\q}(-e_{\q}^{-a}\,z,q) =
  \q\,\varphi_{r, 1 + \frac{s-1}{\q},p,1}(-z^{\q},q^{\q})\,.
\end{equation*}
Thus, we arrive at the following equation for the sought function
$\varphi_{r,s,p,\q}(z,q)$:
\begin{equation}\label{Riemann}
  \sum_{a=0}^{\q-1}\,\varphi_{r,s,p,\q}(e_{\q}^{-a}\,z,q) =
  -\frac{\q\,z^{1 - \q - s}\;                     
    \eta(q^{p\q})^3\,
    \vt{1,1}{q^{r\q}}{q^{p\q}}}{
    \vt{1,1}{(-z)^{\q}}{q^{p\q}}\,
    \vt{1,1}{(-z)^{\q}q^{r\q}}{q^{p\q}}}\,.
\end{equation}
This is to be solved under the condition that the poles of
$\varphi_{r,s,p,\q}(z,q)$ are at $z=-q^{n p}$, $n\in\oN$, and at
$z=-q^{n p - r}$, $n\in\oN$.  Note that $z=-e_{\q}^a\,q^{n p}$ and
$z=-e_{\q}^a\,q^{n p - r}$, with $a=0,\ldots\q-1$ and $n\in\oN$, are
then the complete set of poles of the function of the right-hand side
of~\eqref{Riemann}.

\subsection{Massive admissible
  characters\label{massive-adm-characters}} Characters of the massive
admissible representation $\mAm_{r,s,h,p,\q}$ from
Sec.~{\it\ref{sec:massive-admissible}\/} are read off from the
resolution~\eqref{massive-adm-resolution}. We thus find
\begin{multline}\label{massive-adm-char}
  \charn{\mAm}{r,s,h,p,\q}(z,q)=
  z^h\,\frac{\vt{1,0}{z}{q}}{\eta(q)^3}\,
  q^{\frac{\q}{4p}[(r - \frac{p}{\q}(s-1))^2
    -\frac{\q}{4p}(\frac{p}{\q}h-1)^2]
    - \frac{p\q}{4}}
  \times{}\\
  {}\times\left(
    \vartheta_{1,0}(q^{-p\q - (s-1)p + r\q},q^{2p\q})-
    q^{r(s-1)}
    \vartheta_{1,0}(q^{-p\q - (s-1)p - r\q},q^{2p\q})
  \right),
\end{multline}
which, as we see, has the structure of a massive Verma module
character times a {\it Virasoro\/} character (more precisely, the
ratio of the irreducible Virasoro character to the Virasoro Verma
module character). The relation between the embedding case III$(0)$
and the Virasoro embedding structure has been pointed out
in~\cite{[SSi]}.

\section{The $\protect\tSL2\leftrightarrow\N2$
  correspondence at the level of characters\label{sec:correspondence}}
The construction~\cite{[DvPYZ]} of the $\N2$ algebra, which in fact
determines the simplest Kazama--Suzuki coset~\cite{[KS]} $sl(2)\oplus
u(1)/u(1)$, allows one to completely analyse the {\it
  representations\/} involved in this construction. This was done
in~\cite{[FST]} where the $\tSL2$ and $\N2$ representation categories
were proved to be equivalent up to the respective spectral flows. In
this section, we recall the equivalence and then derive the
consequences for characters.

\subsection{Reminder on the equivalence of
  categories\label{subsec:equivalence}} For simplicity, we avoid
introducing {\it chains\/} of modules and speak instead of
representations {\it modulo the spectral flow\/}; exact definitions
can be found in~\cite{[FST]}. \ We will only need representations with
a \hw{} vector, so all the general statements below are limited to
this case.
\begin{Thm}[\cite{[FST]}]\label{mainthm}
  There exists a functor $\func$ that assigns to every $\tSL2$
  representation with a \hw{} vector $\ketSL{j,t}$ an $\N2$
  representation with the \hw{} vector
  $\bigl|-2j/t,t\bigr\rangle_{\mathrm{top}}$ up to the spectral flow,
  \begin{equation}
    \func(\mC_{j,t})=\mD_{-\frac{2j}{t},t;\bullet}\,.
  \end{equation}
  There exists the inverse functor modulo the spectral flow.  In
  particular, a homomorphism $\mC'_{j',t}\to\mC_{j,t}$ of $\tSL2$
  representations exists if and only if there is a homomorphism of
  $\N2$ representations
  $\mD'_{-\frac{2j'}{2},t;\theta}\to\mD_{-\frac{2j}{t},t}$ with some
  $\theta$ and
  \begin{equation*}
    \func(\mC'_{j',t})=\mD'_{-\frac{2j'}{t},t;\bullet}\,.
  \end{equation*}
\end{Thm}

Underlying this Theorem is the construction of the $\N2$ algebra in
terms of $\tSL2$ and the ghosts~\cite{[DvPYZ],[KS]} and the analysis
of representations initiated in~\cite{[DvPYZ]}: introducing the
fermionic first-order $BC$ system, one can construct the $\N2$
generators as
\begin{alignat}{2}
  \cQ&=CJ^+\,,&
  \cG&=\tfrac{2}{t}BJ^-\,,\label{Gsl}\\
  \cH&=\tfrac{t-2}{t}BC-\tfrac{2}{t}J^0\,,&\qquad
  \cT&=\tfrac{1}{t}(J^+J^-)-
  \tfrac{t-2}{t}B\d C-\tfrac{2}{t}BCJ^0\,,\label{Tsl}
\end{alignat}
where $\cG=\sum_{n\in\oZ}\cG_nz^{-n-2}$,
$\cT=\sum_{n\in\oZ}\cL_nz^{-n-2}$, $\cQ=\sum_{n\in\oZ}\cQ_nz^{-n-1}$,
$\cH=\sum_{n\in\oZ}\cH_nz^{-n-1}$,
$J^{\pm,0}=\sum_{n\in\oZ}J^{\pm,0}_nz^{-n-1}$, and
$B(z)C(w)=\frac{1}{z-w}$.  There also exists an extra current
\begin{equation}\label{Heisenberg}
  I^+=\sqrt{\tfrac{2}{t}}(BC+J^0)
\end{equation}
that commutes with the $\N2$ generators~\eqref{Gsl}--\eqref{Tsl}.

Now, let $\Omega$ be the module generated from the vacuum $\ketGH0$
defined by the \hw{} conditions $C_{\geq1}\ketGH0=B_{\geq0}\ketGH0=0$
and let $\mF^+_\rho$ be the Fock space generated from the \hw{} vector
determined by $I_{\geq1}\,\ket{\rho}^+=0$ and
$I_0\,\ket{\rho}^+=\rho\,\ket{\rho}^+$.

Given an $\tSL2$ representation $\mC_{j,t}$ (e.g., from category
$\cO$), the tensor product $\mC_{j,t}\tensor\Omega$ carries the
representation~\eqref{Gsl}--\eqref{Tsl} of the $\N2$ algebra.  This
representation is highly reducible; decomposing it, we arrive at the
representations $\mD_{-\frac{2j}{t},t;\bullet}=\func(\mC_{j,t})$:
\begin{equation}
  \mC_{j,t;\theta}\tensor\Omega\,{}\approx \bigoplus_{m\in\oZ}\,
  \mD_{-\frac{2j}{t},t;m-\theta}\tensor
  \mF^+_{\sqrt{\frac{2}{t}}(j-\frac{t-2}{2}\theta-m)}\,.
  \label{idspaces}
\end{equation}
This formula, proved in~\cite{[FST]}, is essentially equivalent to the
statement of the theorem.

\subsection{Identities relating $\protect\N2$ and $\protect\tSL2$
  characters\label{sec:identities}} Theorem~\ref{mainthm} has a direct
implication for characters.  Equations~\eqref{Gsl}--\eqref{Heisenberg}
imply the identities
\begin{align}
  T_\mathrm{Sug}+T_\mathrm{gh}&=\cT + T^+\\
  J^0-BC&=-2 \cH + \tfrac{t-4}{\sqrt{2t}}\,I^+\,,
\end{align}
where $T_\mathrm{Sug}$ is the Sugawara \emt{} for the affine~$\SL2$,
$T_\mathrm{gh}=-B\,\d C$ is the ghost \emt, and
\begin{equation}
  T^+=\tfrac{1}{2}\,(I^+)^2-\tfrac{1}{\sqrt{2t}}\d\,I^+.
\end{equation}
We thus have a formal equality
\begin{equation}\label{powers-2nd}
  q^{L_0^\mathrm{Sug}+L_0^\mathrm{gh}}\,z^{J^0_0}\,
  (z y^{-1})^{(BC)_0} =
  q^{\cL_0 + L^+_0}\,y^{-\cH_0 - \sqrt{\frac{2}{t}}I^+_0}
  z^{\sqrt{\frac{t}{2}}\,I^+_0}.
\end{equation}
We now take the traces of both sides over the spaces on the respective
sides of Eq.~\eqref{idspaces}. \ Then, for any two modules related by
the functor, $\mC_{j,t}$ on the $\tSL2$ side
and~$\mD_{-\frac{2j}{t},t}$ on the $\N2$ side, we obtain the following
relation between their characters:
\begin{equation}
  \charsl{\mC}{j,t}(z,q)\,\chi^\mathrm{gh}(z\,y^{-1},q)=
  \sum_{\theta\in\oZ}
  \charn{\mD}{-\frac{2j}{t},t;\theta}(y^{-1},q)\cdot
  \frac{\Bigl(y^{-\sqrt{\frac{2}{t}}}\,z^{\sqrt{\frac{t}{2}}}
    \Bigr)^{\sqrt{\frac{2}{t}}(j-\theta)}
    q^{\frac{(j-\theta)^2 + j-\theta}{t}}}{
    \displaystyle\prod_{m\geq1}(1-q^m)}\,,
\end{equation}
where $1<|z|<|q|^{-1}$.  Recall now that the ghost character is
\begin{equation}
  \chi^{\rm gh}(z,q)=
  \Tr_\Omega(q^{L^{\rm gh}_0} z^{-(BC)_0})
  =\prod_{m=0}^\infty(1+q^m\,z)\,
  \prod_{m=1}^\infty(1+q^m\, z^{-1})=
  q^{-\frac{1}{12}}\,
  \frac{z\,\vartheta_{1,0}(z,q)}{\eta(q)}\,.
\end{equation}
Thus, we have arrived at
\begin{Thm}\label{lemma:identity}
  Let $\mC_{j,t}$ be an $\tSL2$ representation from category $\cO$
  with the \hw{} vector $\ketSL{j,t}$ and let $\func(\mC_{j,t}) =
  \mD_{-\frac{2j}{t},t;\bullet}$. Let $\charsl{\mC}{j,t}(z,q)$ be the
  character of $\mC_{j,t}$ and $\charn{\mD}{-\frac{2j}{t},t}(z, q)$ be
  the character of one of the representations
  $\mD_{-\frac{2j}{t},t;\bullet}$.  Then {\rm(}assuming
  $1<|z|<|q|^{-1}${\rm)},
  \begin{equation}
    \begin{split}\label{identity-new}
      \charsl{\mC}{j,t}(z,q)\,\vt{1,0}{\tfrac{z}{y}}{q}&=
      \sum_{\theta\in\oZ}
      \charn{\mD}{-\frac{2j}{t},t;\theta}(y^{-1}, q)\,
      y^{-\frac{2}{t}(j-\theta)}\,z^{j-\theta}\,
      q^{\frac{(j-\theta)^2+j-\theta}{t} + \frac{1}{8}}\\
      &= \sum_{\theta\in\oZ}
      \charn{\mD}{-\frac{2j}{t},t}(y^{-1}\,q^{-\theta}, q)\,
      y^{\theta-\frac{2}{t}j}\,z^{j-\theta}\,
      q^{\frac{\theta^2-\theta}{2} +
        \frac{j^2 + j - 2 j \theta}{t} + \frac{1}{8}}.
    \end{split}
  \end{equation}
\end{Thm}
\begin{Rem}
  Taking a {\it twisted\/} module on the $\tSL2$ side, we can proceed
  in a similar way and arrive at
  \begin{multline}\label{identity-new-twist}
      \charsl{\mC}{j,t;\mu}(z,q)\,\vt{1,0}{\tfrac{z}{y}}{q}=\\
      {}=z^{-\frac{t}{2}\mu}y^\mu q^{\frac{t}{4}\mu^2-\mu(j+\half)}\,
      \sum_{\theta\in\oZ}
      \charn{\mD}{-\frac{2j}{t},t}(y^{-1}\,q^{-\theta}, q)\,
      y^{\theta-\frac{2}{t}j}\,z^{j-\theta}\,
      q^{\frac{\theta^2-\theta}{2} + \mu\theta +
        \frac{j^2 + j - 2 j \theta}{t} + \frac{1}{8}},
  \end{multline}
  where $|q|^{\mu}<|z|<|q|^{\mu-1}$.
\end{Rem}

Equation~\eqref{identity-new} can be viewed as a ``generalised
branching rule'' (or, if read from right to left, a ``sum rule''),
termed ``generalised'' because it contains an {\it infinite\/} number
of terms on the right-hand side. The most interesting cases are when
the characters on the right-hand side are quasi-periodic and, thus,
the summation contains only a finite number of terms. This will be
considered in Sec.~{\it\ref{sec:unitary-branching}}, where we obtain
true branching relations.  However, even with the nonperiodic
Verma-module characters, Eq.~\eqref{identity-new} becomes the
$\tSSL21$ denominator identity, see~\eqref{truly}.

It should be kept in mind that $\func$ has the inverse functor, and
therefore, {\it every\/} $\N2$ representation admits a ``sum rule''
yielding the character of the corresponding $\tSL2$ representation
after taking the trace over the spectral flow transforms, as
in~\eqref{identity-new}.

\subsection{An integral representation for $\N2$
  characters\label{sec:integral-general}}
Relation~\eqref{identity-new} implies an integral representation for
the $\N2$ characters:
\begin{equation}\label{integral-general}
  \charn{\mD}{h,t}(y, q)\,=
  y^h\,q^{-\frac{h}{4}(ht-2)-\frac{1}{8}}\frac{1}{2\pi i}\oint\! dz\,
  z^{-\frac{h t}{2}-1}\,
  \charsl{\mC}{-\frac{h t}{2},t}(z,q)\,\vt{1,0}{z\,y}{q}\,.
\end{equation}
As before, $\mD_{h,t}$ and $\mC_{j,t;\bullet}$ are some $\N2$ and
$\tSL2$ representations, respectively, related by the functor. The
integration contour encompasses the origin counterclockwise and lies
inside the annulus determined by the twist of the chosen $\mC$
representation.

\subsection{Examples} Let us see how the general formula of
Theorem~\ref{lemma:identity} rewrites for particular representations.

\subsubsection{Trivial representations} For the trivial
representations ($t=2$, $j=0$, $\charsl{\mC}{0,2; \mu}(z,q)=1$, and
$\charn{\mD}{0,2}(z,q)=1$), Eq.~\eqref{identity-new} becomes simply
the definition of $\vt{1,0}{\frac{z}{y}}{q}$.

\subsubsection{Topological Verma modules} Taking the $\N2$
representations to be the twisted topological Verma modules, and the
$\tSL2$ representation, accordingly, the Verma module, we see that
Eq.~\eqref{identity-new} becomes the ``{\it truly remarkable
  identity\/}'' from~\cite[Example~4.1]{[KW]}:
\begin{align}
  \frac{\vartheta_{1,0}(zy^{-1},q)\,\eta(q)^3}{
    \vartheta_{1,0}(y^{-1},q)\vartheta_{1,1}(z,q)} &=
  \sum\limits_{\theta\in\oZ}\frac{z^{-\theta}}{ 1+yq^\theta}\,,
  \qquad 1<|z|<|q|^{-1},\label{truly}\\
  &=\Bigl(\sum_{\substack{m\geq n\\
      k\geq0}}- \sum_{\substack{m\leq n-1\\ k\leq-1}}\Bigr)
  z^{-m}\,(-1)^k\,y^{k}\,q^{m k}\,, \quad
  \begin{gathered}
    1<|z|<|q|^{-1}, \quad\\
    |q|^{-n+1}<|y|<|q|^{-n}.
  \end{gathered}\label{truly-2}
\end{align}
Taking the $\tSL2$ Verma module with an arbitrary twist, we have,
similarly,
\begin{equation}\label{truly-mu}
  (-1)^\mu y^{-\mu}\frac{\vartheta_{1,0}(zy^{-1},q)\,\eta(q)^3}{
    \vartheta_{1,0}(y^{-1},q)\vartheta_{1,1}(z,q)} =
  \sum\limits_{m\in\oZ}\frac{z^{-m}q^{\mu m}}{ 1+yq^m}\,,
  \qquad |q|^\mu<|z|<|q|^{\mu-1}\,.
\end{equation}
As a particular case, we recover the identity that was written out
already in~\cite{[FST]}:
\begin{equation}\label{identity1}
  \frac{\vartheta_{1,0}(z,
    q)}{\vartheta_{1,1}(z,q)}= \frac{\vartheta_{1,0}(z^2, q)}{
    \eta(q)^3}\sum_{\theta\in\oZ}
  \frac{z^{-\theta+1}}{1+z^2\,q^\theta}\,,\qquad 1<|z|<|q|^{-1}.
\end{equation}

\subsubsection{Admissible representations: resummation and the
  integral formula\label{sec:admissible-branching}} For the admissible
representation characters given by~\eqref{admissible-N2}
and~\eqref{admissible-sl2}, Eq.~\eqref{identity-new} becomes
\begin{multline}\label{identity3}
  \sum_{\theta\in\oZ}z^{-\theta}
  \varphi_{r,s,p,\q}(y^{-1}\,q^{-\theta}, q) = q^{-\frac{p\q}{4}}
  \frac{\vt{1,0}{z\,y^{-1}}{q}\,\eta(q)^3}{
    \vt{1,1}{z}{q}\,\vt{1,0}{y^{-1}}{q}} \left(
    \vt{1,0}{z^p\,q^{-p\q - p(s-1) + \q r}}{q^{2p\q}}\right.\\
  \left.{}-q^{r(s-1)}\,z^{-r} \vt{1,0}{z^p\,q^{-p\q - p(s-1) - \q
        r}}{q^{2p\q}} \right),
\end{multline}
where $1<|z|<|q|^{-1}$.  This can also be obtained from the previous
identities by first deriving
\begin{multline}\label{phi-identity2}
  \sum_{\theta\in\oZ}z^{-\theta}
  \varphi_{r,s,p,\q}(y^{-1}\,q^{-\theta}, q) =\\
  {}=q^{-\frac{p\q}{4}}
  \sum_{\theta\in\oZ}\frac{z^{-\theta}}{1 + y\,q^{\theta}}\cdot
  \left(
    \vt{1,0}{z^p\,q^{-p\q - p(s-1) + \q r}}{q^{2p\q}}
    -q^{r(s-1)}\,z^{-r}
    \vt{1,0}{z^p\,q^{-p\q - p(s-1) - \q r}}{q^{2p\q}}
  \right)
\end{multline}
and then combining this with~\eqref{truly}.

\medskip

Recall that the function $\varphi_{r,s,p,\q}(z,q)$ defined
in~\eqref{varphi} cannot be algebraically expressed through theta
functions (unless $\q=1$), since it is not quasi-periodic.  On the
other hand, there is an {\it integral\/} representation for
$\varphi_{r,s,p,\q}(z,q)$ with the integrand given by a combination of
theta functions. The integral formula for the admissible $\N2$
characters takes the form of the following representation for
$\varphi_{r,s,p,\q}$:
\begin{multline}\label{integral-admissible}
  \frac{\vt{1,0}{y}{q}}{\eta(q)^3}\,
  \varphi_{r,s,p,\q}(y,q)=
  \frac{1}{2\pi i}\oint_{\cC_{\q}(0)}
  \frac{d z}{z}\,\vt{1,0}{zy}{q}\,
  \frac{q^{-\frac{1}{4}p\q}}{
    \vartheta_{1,1}(z, q)}\times{}\\
  {}\times\left(
    \vartheta_{1,0}(z^p\,q^{-p\q  +  r \q  - (s-1) p}, q^{2p\q})
    -z^{-r}\,q^{r(s-1)}\,
    \vartheta_{1,0}(z^p\,q^{-p\q  -  r \q  - (s-1) p}, q^{2p\q})
  \right),
\end{multline}
where we use the notation $\cC_{\q}(n)$ for the integration contour
that surrounds the origin counterclockwise and lies in the annulus
\begin{equation*}
  |q|^{-2\q n} < |z| < |q|^{-2\q n - 1}\,.
\end{equation*}
In the present case of $n=0$, the integral is given by the sum over
the poles inside the unit circle. Now, under the spectral flow
transform with $\theta=p n$, the integration contour changes as
$\cC_{\q}(0)\to\cC_{\q}(n)$, which is described by adding or
subtracting the corresponding poles described in
Lemma~\ref{admissible-poles}.
We parametrise these~as $z=q^n$, $n=-2\q \ell - a - 1$, $\ell\in\oZ$.
Then
\begin{multline}
  \res_{z=q^{-2\q \ell - a - 1}} \biggl(z^{-1}\vt{1,0}{zy^{-1}}{q}\,
  \frac{q^{-\frac{1}{4}p\q}}{
    \vartheta_{1,1}(z, q)}\times{}\\
  {}\times\left( \vartheta_{1,0}(z^p\,q^{-p\q + r \q - (s-1) p},
    q^{2p\q}) -z^{-r}\,q^{r(s-1)}\, \vartheta_{1,0}(z^p\,q^{-p\q - r
      \q - (s-1) p}, q^{2p\q})
  \right)\biggr)={}\displaybreak[0]\\
  {}=\frac{\vt{1,0}{y^{-1}}{q}\,}{\eta(q)^3}\, (-1)^{a+1}\, y^{-2\q
    \ell - a -1}\,q^{-p\q/4}\,q^{-p\q \ell^2 - \ell p a - \ell p s +
    \ell\q r}
  \times{}\\
  {}\times\Bigl( \vt{1,0}{q^{-p a - p s - p\q+r\q}}{q^{2p\q}} - q^{r(a
    + s)}\vt{1,0}{q^{-p a - p s - p\q - r\q}}{q^{2p\q}}\Bigr),
\end{multline}
which allows us to rederive
Eqs.~\eqref{phi-identity}--\eqref{phi-identity2-} that were obtained
directly from~\eqref{varphi}, and, as a by-product, explains why the
terms violating quasiperiodicity of the admissible $\N2$ characters
are precisely the residues of the $\tSL2$ admissible characters.

\subsubsection{Unitary representations: branching, or sum rule,
  relations\label{sec:unitary-branching}} This case is of some special
interest from our present point of view because the $\N2$ characters
are quasiperiodic, see~\eqref{periodicity-N2-full}. \ This allows us
to rewrite~\eqref{identity-new} in the form with only a finite number
of terms on the right-hand side. The summation goes over the orbit of
the $\N2$ spectral flow on the unitary representations:
\begin{multline}\label{unitary-identity}
  \charsl{\mL}{r,p}(z,q)\,\vt{1,0}{\tfrac{z}{y}}{q}={}\\
  = q^{\frac{r^2-1}{4p} - \frac{p}{4} + \frac{1}{8}}\,
  y^{- \frac{r-1}{p}}\,z^{\frac{r-1}{2}}
  \sum_{a=0}^{p-1}
  \charn{\mK}{r,p}(y^{-1}\,q^{-a}, q)\,
  y^{a}\,z^{-a}\,
  q^{\frac{a^2-a}{2} - \frac{r-1}{p}a }\;
  \vt{1,0}{z^p\,y^{-2}\,q^{-p+r-2a}}{q^{2p}}\,.
\end{multline}
Explicitly substituting the $\tSL2$ and $\N2$ characters given by
\eqref{integrable-sl2} and \eqref{N2-unitary}, respectively, we obtain
\begin{multline}\label{identity-unitary}
  \frac{\vartheta_{1,0}(z^p\,q^{-p  +  r}, q^{2p}) -
    z^{-r}
    \vartheta_{1,0}(z^p\,q^{-p  -  r}, q^{2p})}{
    \vartheta_{1,1}(z, q)}\,
  \vt{1,0}{zy}{q}={}\\
  {}= -\frac{\eta(q^p)^3}{\eta(q)^3}\,              
  \vt{1,1}{q^r}{q^p}\,\vt{1,0}{y}{q}
  \sum_{a=0}^{p-1}
  \frac{z^{-a}q^a\,y^{-1}\vt{1,0}{z^py^{2}q^{r-p-2a}}{q^{2p}}}{
    \vt{1,0}{yq^{-a}}{q^p}\,\vt{1,0}{yq^{r-a}}{q^p}
    }\,.
\end{multline}

In the special case where $p=2r$, the theta-function identity contains
only $r$ terms on the right-hand side,
\begin{equation}\label{unitary-identity-special}
  \frac{\vt{1,1}{z^r}{q^r}}{\vt{1,1}{z}{q}}\,
  \vt{1,0}{z\,y}{q} = \frac{\eta(q^r)^3}{\eta(q)^3}\,  
  \vt{1,0}{y}{q}\,
  \sum_{a=0}^{r-1}
  \frac{\vt{1,0}{z^r\,y\,q^{-a}}{q^r}}{
    \vt{1,0}{y\,q^{-a}}{q^r}}\,z^{-a}\,,
\end{equation}
where we have used Eqs.~\eqref{p=2r-sl2} and~\eqref{p=2r-N2} for the
$\tSL2$ and $\N2$ characters, respectively.

\section{Concluding remarks} In this paper, we have evaluated the
characters of admissible $\N2$ representations, making the derivation
parallel to the affine-$\SL2$ case and explicitly deriving the exact
relations between the $\N2$ and $\tSL2$ characters.  An important tool
in our analysis of characters has been the spectral flow transform of
the $\tSL2$ and $\N2$ algebras.

\smallskip

We have constructed different resolutions of the admissible $\N2$
representations.  Although the embedding structure~\cite{[SSi]} of
massive $\N2$ Verma modules is more complicated than in the familiar
case of $\tSL2$ Verma modules, it is still not very difficult to
rewrite the embedding diagrams as the resolutions consisting of
massive Verma modules (we have considered in detail the most involved
case III$^0_+(2,{-}{+})$ from~\cite{[SSi]}, others can be dealt with
along the same lines). In the massive resolutions, the number of
modules grows from term to term.

\smallskip

There also exists a simpler resolution in terms of twisted topological
Verma modules, whose structure is in fact equivalent to the well-known
BGG resolution for irreducible $\tSL2$ representations (the character
formulae following from the massive resolution have been shown to
agree with those derived from the BGG resolution).  This has a deep
reason that rests in the equivalence of $\tSL2$ and $\N2$
representation categories up to the spectral flows~\cite{[FST]}. \ The
equivalence can also be viewed as the ``reason'' why the earlier
proposal~\cite{[M]} for the unitary $\N2$ characters is correct
despite a number of subtleties that have not been explicitly taken
into account in~\cite{[M]}. \ A detailed analysis of the structure of
the $\N2$ Verma modules~\cite{[ST4],[SSi]} has allowed us to find the
characters in a considerably more general case of the admissible $\N2$
representations. Not being quasiperiodic under the spectral flow,
these characters cannot be algebraically expressed through the
theta-functions. On the other hand, we have found an integral
representation for these characters through the respective (hence,
admissible) $\tSL2$ characters (which {\it are\/} expressed in terms
of the theta-functions).

\smallskip

We have shown that a certain sum of the $\N2$ characters over the
orbit of the $\N2$ spectral flow provides a kind of ``harmonic
decomposition'' of an $\tSL2$ character. This identity relating the
$\N2$ and $\tSL2$ characters can be considered as the exact
``branching rule'' that holds for {\it any\/} representation, not
necessarily the unitary ones, although it is only for the unitary
representations that the identity involves a finite number of terms on
the $\N2$-side. The resulting theta-function identities are,
therefore, also a consequence of the equivalence of categories.

\smallskip

The equivalence of categories, which we have used for the topological
Verma modules and the irreducible representations of the corresponding
\hw{} type, can also be formulated for the massive $\N2$ Verma modules
(relating them to the relaxed Verma modules over $\tSL2$~\cite{[FST]})
and for the massive-admissible representations
\eqref{massive-adm-resolution}.
The corresponding character identities can easily be derived, however
they are not very interesting.

\smallskip

The analysis of this paper can be viewed from the broader perspective
of coset models; a part of what we have done was to investigate the
$\N2$ algebra pretending to know only that this is $\tSL2\oplus
u(1)/u(1)$. As regards more general cosets, very suggestive pieces of
the picture that we have completely described in the $\N2$ case can be
found in~\cite{[PT]} in a more complicated case of $\N4$ algebras (in
fact, one can ask about different ``sumrules'' for characters whether
they follow from some relations between {\it representations\/}).
Note also that both the $\N2$ and $\tSL2$ algebras are closely related
to the affine Lie superalgebra $\tSSL21$,
see~\cite{[S-sl21],[S-sl21sing]} and references therein.  Thus, it
would be interesting to apply the present approach to the results
of~\cite{[BHT]}, especially because the denominator $\tSSL21$ identity
is already the particular case of the formula expressing the
$\N2\leftrightarrow\tSL2$ equivalence obtained by taking the Verma
modules in the general identity~\eqref{identity-new}.

\bigskip

\noindent{\it Acknowledgements}. We are grateful to V.~Ya.~Fainberg,
A.~V.~Odessky, and M.~A.~Soloviev for useful discussions.  AMS and IYT
were encouraged by very stimulating discussions with F.~Malikov. IYT
wishes to thank A.~A.~Kirillov and I.~Paramonova for a discussion.
AMS is grateful to P.~H.~Damgaard for kind hospitality at the Niels
Bohr Institute, where a part of this paper was written, and to
I.~Shchepochkina for a discussion. VAS is thankful to A.~V.~Gurevich
for his kind attention to her work. This work was supported in part by
the RFBR Grant 98-01-01155.

\appendix

\section{Theta-function conventions\label{app:theta}} Since we deal
with Verma-module characters along with the characters of various
irreducible representations, we prefer using Jacobi theta-functions
(rather than the higher-level theta functions) for all of the
characters that can be expressed through theta-functions.  We thus
introduce the Jacobi theta functions
\begin{alignat}{2}
  \vartheta_{1,1}(z, q) &=
  q^{1/8} \sum_{m\in\oZ}(-1)^m q^{\half(m^2 - m)} z^{-m}
  {}&=
  q^{1/8}\prod_{m\geq0}(1 - z^{-1} q^m)
  \prod_{m\geq1}(1 - z q^m)\prod_{m\geq1}(1 - q^m)\,,\displaybreak[2]\\
  \vt{1,0}{z}{q}&=
  q^{1/8}\sum_{m\in\oZ}^{}q^{\half(m^2 - m)} z^{-m}
  {}&=q^{1/8}
  \prod_{m\geq0}(1+z^{-1}q^m)\prod_{m\geq1}(1+z q^m)
  \prod_{m\geq1}(1-q^m)\,.
\end{alignat}
Then, for $\theta\in\oZ$,
\begin{align}
  \vartheta_{1,1}(z\,q^\theta, q) &=
  (-1)^\theta\,q^{-\half(\theta^2+\theta)}\,
  z^{-\theta}\vartheta_{1,1}(z,
  q)\,,\displaybreak[2]\\
  \vartheta_{1,0}(z\,q^\theta, q) &=
  q^{-\half(\theta^2+\theta)}\,z^{-\theta}\vartheta_{1,0}(z, q)\,.
\end{align}

We also use the Dedekind eta-function
\begin{equation}
  \eta(q)  =q^{\frac{1}{24}}
  \sum_{m=0}^\infty (-1)^m q^{\half(3m^2 + m)}
  =q^{\frac{1}{24}}
  \prod\limits_{m=1}^\infty(1-q^{m})\,.
\end{equation}

The following useful identities are elementary to prove (the prime
means $\d/\d z$):
\begin{align}
  \vt{1,0}{z}{q}&=\frac{q^{\frac{3}{4}}\,\eta(q)}{
    \eta(q^2)^2}\,z\,\vt{1,0}{z}{q^2}\,
  \vt{1,0}{zq}{q^2}\,,\label{n2trivial}\displaybreak[0]\\
  \vt{1,1}{z}{q}&=
  q^{-\frac{3}{8}}
  \left(
    \vartheta_{1,0}(z^2\,q^{-1}, q^4)-
    z^{-1}
    \vartheta_{1,0}(z^2\,q^{-3}, q^4)
  \right)\,,\label{sl2trivial}\displaybreak[0]\\
  \left.\vt{1,0}{z}{q}'\right|_{z=-q^n}&=
  (-1)^{n+1}\,\eta(q)^3\,q^{-\frac{n^2}{2} - \frac{3n}{2}},\quad
  n\in\oZ\,,\label{theta-prime}\displaybreak[0]\\
    \left.\vt{1,1}{z}{q}'\right|_{z=q^n}&=
  (-1)^{n}\,\eta(q)^3\,q^{-\frac{n^2}{2} - \frac{3n}{2}},\quad
  n\in\oZ\,,\label{theta1-prime}\displaybreak[0]\\
  \vartheta_{1,1}(q,q^2)&=-q^{-\frac{3}{4}}\,
  \frac{\eta(q)^2}{\eta(q^2)}\,,\qquad
  \vartheta_{1,0}(1,q)=2\,\frac{\eta(q^2)^2}{\eta(q)}\,.\label{eta-eta}
\end{align}


\begin{thebibliography}{33}

\bibitem{[Ade]}M.~Ademollo, 
  L.~Brink, A.~D'Adda, R.~D'Aura, E.~Napolitano, S.~Sciuto,
  E.~Del~Giudice, P.~Di~Vecchia, S.~Ferrara, F.~Gliozzi, R.~Musto,
  R.~Pettorino and J.~Schwarz, {\it Dual String With U(1) Color
    Symmetry}, \NPB111 (1976) 77;\\
  M.~Ademollo, 
  L.~Brink, A.~D'Adda, R.~D'Aura, E.~Napolitano, S.~Sciuto,
  E.~Del~Giudice, P.~Di~Vecchia, S.~Ferrara, F.~Gliozzi, R.~Musto and
  R.~Pettorino, {\it Dual String Models With Nonabelian Color and
    Flavor Symmetries}, \NPB114 (1976) 297.

\bibitem{[FST]}B.L.~Feigin, A.M.~Semikhatov, and I.Yu.~Tipunin,
  {\it Equivalence between Chain Categories of Representations of
    Affine $\SL2$ and $N=2$ Superconformal Algebras}, hep-th/9701043,
  J. Math. Phys. (1998).

\bibitem{[KW0]} V.~G.~Ka\v c and M.~Wakimoto, {\it Modular Invariant
    Representations of Infinite-Dimensional Lie Algebras and
    Superalgebras}, {\sl Proc. of the National Academy of Sciences
    USA}, 85 (1988) 4956.

\bibitem{[Dob]}V.K.~Dobrev, {\it Characters of Unitarizable
    Highest Weight Modules over the $\N2$ Superconformal Algebra},
  \PLB186 (1987) 43.

\bibitem{[M]}Y.~Matsuo, {\it Character Formula of $C<1$ Unitary
    Representation of $\N2$ Superconformal ALgebra}, Prog.\ Theor.\
  Phys. 77 (1987) 793--797.

\bibitem{[RY]} F.~Ravanini and S.-K.~Yang, {\it Modular Invariance
    in $\N2$ Superconformal Theories}, \PLB195 (1987) 202--208.

\bibitem{[Kir]}E.B.~Kiritsis, {\it Character Formulae and
    the Structure of the Representations of the $\N1$, $\N2$
    Superconformal Algebras}, \IJMPA 3 (1988) 1871--1906.

\bibitem{[KW]} V.G.~Ka\v c and M.~Wakimoto, {\it Integrable Highest
    Weight Modules over Affine Superalgebras and Number Theory},
  hep-th/9407057.

\bibitem{[Doerr2]}M.~D\"orrzapf, {\it Analytic Expressions for
    Singular Vectors of the $\N2$ Superconformal Algebra}, Commun.\
  Math.\ Phys.\ 180 (1996) 195.

\bibitem{[ST4]}A.M.~Semikhatov and I.Yu.~Tipunin, {\it The Structure
    of Verma Modules over the $N=2$ Superconformal Algebra\/},
  Commun.\ Math.\ Phys., 195 (1998) 129-173.

\bibitem{[SSi]} A.M.~Semikhatov and V.A.~Sirota, {\it Embedding
    Diagrams of $\N2$ and Relaxed-$\tSL2$ Verma Modules},
  hep-th/9712102.

\bibitem{[RCW]}A.~Rocha-Caridi and N.R.~Wallach, {\it Highest Weight
    Modules over Graded Lie Algebras: Resolutions, Filtrations, and
    Character Formulas}, Trans. Amer. Math. Soc. 277 (1983) 133--162.

\bibitem{[Mal]}F.~Malikov, {\it Verma Modules over Rank-2 Ka\v
    c--Moody Algebras}, Algebra i Analiz 2 No.~2 (1990) 65.



\bibitem{[JL]} K.~Junemann and O.~Lechtenfeld, {\it Chiral BRST
    Cohomology of N=2 Strings at Arbitrary Ghost and Picture Number},
  hep-th/9712182.


\bibitem{[ST3]}A.M.~Semikhatov and I.Yu.~Tipunin, {\it All Singular
    Vectors of the $N=2$ Superconformal Algebra via the Algebraic
    Continuation Approach\/}, hep-th/9604176.

\bibitem{[TheBook]} V.G.~Ka\v c {\sl Infinite Dimensional Lie
    Algebras\/}, Cambridge University Press, 1990.

\bibitem{[BH]}K.~Bardak\c ci and M.B.~Halpern, {\it New Dual Quark
    Models}, Phys.\ Rev.\ D3 (1971) 2493--2506.

\bibitem{[SS]}A.~Schwimmer and N.~Seiberg, {\it Comments on the
    $N=2$, $N=3$, $N=4$ Superconformal Algebras in Two-Dimensions\/},
  \PLB184 (1987) 191.

\bibitem{[W-Gep]}D.~Gepner and E.~Witten, \NPB278 (1986) 493.

\bibitem{[DvPYZ]} P.~Di~Vecchia, J.L.~Petersen, M.~Yu, and H.B.~Zheng,
  {\it Explicit Construction of Unitary Representations of the $\N2$
    Superconformal Algebra}, \PLB174, (1986) 280--284.


\bibitem{[KS]}Y.~Kazama and H.~Suzuki, {\it New $N=2$ Superconformal
    Field Theories and Superstring compactification}, \NPB321 (1989)
  232.

\bibitem{[LVW]}W.~Lerche, C.~Vafa, and N.P.~Warner, {\it Chiral
    Rings in N=2 Superconformal Theories\/}, \NPB324 (1989) 427.

\bibitem{[BFK]}W.~Boucher, D.~Friedan, and A.~Kent, {\it
    Determinant Formulae and Unitarity For the $N=2$ Superconformal
    Algebras in Two-Dimensions or Exact Results on String
    Compactification\/}, \PLB172 (1986) 316--322.
  
\bibitem{[D-emb]} M.~D\"{o}rrzapf, {\it The Embedding Structure of
    Unitary $N=2$ Minimal Models}, hep-th/9712165.

\bibitem{[FST-semi]} B.L.~Feigin, A.M.~Semikhatov, and I.Yu.~Tipunin,
  {\it A Semi-Infinite Realization of Unitary Representations of the
    $\N2$ Superconformal Algebra}, to appear.

\bibitem{[FSt]} B.L.~Feigin and A.V.~Stoianovsky, {\it Functional
    Models of Representations of Current Algebras and Semi-Infinite
    Schubert Cells}, Funk. An. i ego Prilozh., 28(1) (1994) 68.

\bibitem{[BT]}P.~Bowcock and A.~Taormina, {\it Representation Theory
    of the Affine Lie Superalgebra $sl(2|1)$ at Fractional Level\/},
  Commun. Math. Phys. 185 (1997) 467-493.



\bibitem{[BGG]}I.~Bernshtein, I.~Gelfand, and S.~Gelfand, Funk.
  An. Prilozh. 10 (1976) 1.

\bibitem{[PT]} J.L.~Petersen and A.~Taormina, {\it Coset
    Construction and Character Sumrules for the Doubly Extended $N=4$
    Superconformal Algebras}, \NPB398 (1993) 459.

\bibitem{[S-sl21]}A.M.~Semikhatov, {\it The Non-Critical $N=2$
    String is an $\SSL21$ Theory\/}, \NPB478 (1996) 209--234.

\bibitem{[S-sl21sing]} A.M.~Semikhatov, {\it Verma Modules,
    Extremal Vectors, and Singular Vectors on the Non-Critical $\N2$
    String Worldsheet}, hep-th/9610084.

\bibitem{[BHT]} P.~Bowcock, M.~Hayes, and A.~Taormina, {\it Characters
    of admissible representations of the affine Lie
    superalgebra $\hat{sl}(2|1;\oC)_k$}, DTP/97/19.\\
  M.~Hayes and A.~Taormina, {\it Admissible $\hat{sl}(2|1;\mathbb
    C)_k$ Characters and Parafermions}, hep-th/9803022.

\end{thebibliography}
\end{document}